\documentclass[]{aa} 

\usepackage{graphicx}
\usepackage{txfonts}
\usepackage[]{hyperref}  
\usepackage{threeparttable}

\hypersetup{
  colorlinks=true,  
  urlcolor=blue,    
  linkcolor=red,
  citecolor=blue     
}

\usepackage{appendix}
\usepackage{etoolbox}
\makeatletter
\patchcmd\@combinedblfloats{\box\@outputbox}{\unvbox\@outputbox}{%
    \errmessage{\noexpand\@combinedblfloats could not be patched}%
}
\makeatother
\begin{document}

   \title{On the intensity ratio variation of the \ion{Si}{IV} 1394/1403 \AA\ lines during solar flares}
   \subtitle{Resonance scattering and opacity effects}

   \author{Hao-Cheng Yu
          \inst{1,2}
          \and
          J. Hong\inst{1,2}
          \and
          M. D. Ding\inst{1,2}
          }

   \institute{School of Astronomy and Space Science, Nanjing University, 163 Xianlin Road, Nanjing 210023, PR China\\
              \email{jiehong@nju.edu.cn}
         \and
             Key Laboratory for Modern Astronomy and Astrophysics (Nanjing University), Ministry of Education, Nanjing 210023, 163 Xianlin Road, PR China
             }

   \date{}

 
  \abstract
   {The \ion{Si}{IV} lines at 1394 \AA\ and 1403 \AA\ form in the solar atmosphere at a temperature of $\sim10^{4.8}$ K. They are usually considered optically thin, but their opacity can be enhanced during solar flares. Traditionally, the intensity ratio of these lines are used as an indicator of the optical thickness. However, observations have shown a wavelength-dependent intensity ratio profile $r(\Delta\lambda)$ of the the 1394 \AA\ to 1403 \AA\ lines.}
   {We aim to study the variation of the intensity ratio profile in solar flares and the physical reasons behind it.}
   {The \ion{Si}{IV} lines and their intensity ratio profiles are calculated from the one-dimensional radiative hydrodynamics flare model with non-thermal electron heating.}
   {During flares, $r(\Delta\lambda)$ is smaller than 2 at the line core but larger than 2 at the line wings. We attribute the deviation of the ratio from 2 to two effects: the resonance scattering effect and the opacity effect. Resonance scattering increases the population ratio of the upper levels of the two lines, and as a result, increases $r(\Delta\lambda)$, in all wavelengths. The opacity effect decreases $r(\Delta\lambda)$, especially at the line core where the opacity is larger. These two effects compete with each other and cause the U-shape of $r(\Delta\lambda)$.}
   {}

   \keywords{sun: chromosphere --
                line: profiles --
                radiative transfer --
                opacity
               }

   \maketitle
%

\section{Introduction}

The transition region (TR) is a thin layer between the chromosphere and the corona, where the physical quantities change dramatically in the Sun. The temperature  could change by several orders of magnitude, and a number of emission lines  form in the TR, such as the \ion{Si}{IV}, \ion{C}{IV}, \ion{O}{IV}, and \ion{Ne}{VII} lines, which could reveal the dynamic properties of local plasma. The \ion{Si}{IV} resonance lines at 1394 \AA\ and 1403 \AA\ are two typical TR lines with a formation temperature of $\sim10^{4.8}$ K. It is generally accepted that photons generated in the TR are able to escape from the solar surface without absorption or scattering; that is, these lines are usually considered optically thin. Under the optically thin assumption, \cite{1999A&A...351L..23M} indicated that, for pairs of resonance lines such as \ion{Si}{IV} $1394/1403$ \AA\ and \ion{C}{IV} $1548/1551$ \AA\ which share the same lower level but different upper levels, the ratio of the line intensities is exactly the ratio of the oscillator strengths of the two lines. This ratio is 0.52:0.26=2:1 for the \ion{Si}{IV} 1394/1403 \AA\ lines \citep{1993PhLA..173..407M}. Usually, a decrease from 2 implies that there is an opacity effect along the line of sight. 

During heating events like transient brightenings in the transition region or solar flares, the atmosphere is heated to a high temperature where the \ion{Si}{IV} lines could be significantly enhanced. Recent works suggested that the optically thin assumption of the \ion{Si}{IV} lines does not hold in some cases. \cite{2015ApJ...811...48Y} found the self-absorption effect on both \ion{Si}{IV} lines for the first time. They found that the intensity ratio is reduced to 1.7$\sim$2. The same  effect has also been reported in \cite{2017ApJ...845...16N}. \cite{2020ApJ...894..128T} studied the distribution of intensity ratios in active regions. They concluded that there are a considerable number of pixels where the intensity ratio deviates from 2, especially in the early phases of flares and in parts of flare ribbons. Recent radiative hydrodynamic simulations indicated that the two \ion{Si}{IV} lines might  not be optically thin in flare conditions \citep{2019ApJ...871...23K}, where the intensity ratio varies from 1.8 to 2.3. 

It should be pointed out that the intensity ratio could also be larger than 2 in many observations \citep{2020ApJ...894..128T,2022zhou}, where the opacity effect fails to explain. Previous studies attributed this to the result of resonance scattering, which implies that a stronger resonance scattering yields a larger intensity ratio \citep{2018A&A...619A..64G,2013A&A...550A..16G}.

The above mentioned intensity ratio usually refers to the ratio of wavelength-integrated intensity $R=\int I_{1394 \AA}(\lambda)\mathrm{d}\lambda/\int I_{1403 \AA}(\lambda)\mathrm{d}\lambda$ as adopted in previous studies. However, considering that the opacity at the line core is much larger than that at the line wings, the intensity ratio may also vary from the line core to the line wings. \cite{2022zhou} focused on the wavelength-dependent intensity ratio profile $r(\Delta\lambda)=I_{1394 \AA}(\Delta\lambda)/I_{1403 \AA}(\Delta\lambda)$. They found that at the flare ribbons, $r(\Delta\lambda)$ is less than 2 at the line core and larger than 2 at the line wings, while the ratio of the integrated intensity is still close to 2. It is suggested that the intensity ratio profile $r(\Delta\lambda)$ serves as a better proxy of the line opacity. Thus, the variation of $r(\Delta\lambda)$ along wavelength deserves further investigation.

In this paper, we study the intensity ratio profile $r(\Delta\lambda)$ of the \ion{Si}{IV} resonance lines in one-dimensional radiative hydrodynamic flare models. We examine the variation of the intensity ratio along wavelength and explore possible physical mechanisms. The flare models and calculation of the \ion{Si}{IV} lines are briefly  described in Section~\ref{sec2}. The spectral line profiles and intensity ratio profiles $r(\Delta\lambda)$ are shown in Section~\ref{sec3}. Reasons for the variation of $r(\Delta\lambda)$ are discussed in Section~\ref{sec4}. Finally we make conclusions in Section~\ref{sec5}.

\section{Method}
\label{sec2}
The one-dimensional radiative hydrodynamics code \verb"RADYN" was first used to study shocks in the chromosphere \citep{1992ApJ...397L..59C,1997ApJ...481..500C,1995ApJ...440L..29C}, but now it is more often employed for the response of the solar atmosphere during flares \citep{2015ApJ...809..104A}. The conservation equations of mass, momentum, energy and charge coupled with atom level population equations and radiative transfer equations are solved in a 1D plane-parallel atmosphere, on an adaptive grid \citep{1987Dorfi}. The initial atmosphere is based on the VAL3C model, with a 10 Mm semicircular flare loop. The loop-top temperature is 10 MK. To simulate the heating by flares, a high energy non-thermal electron beam is injected from the loop top, and then moves downward along the flare loop. A set of three parameters describes the power-law energy distribution of non-thermal electrons: the cut-off energy $E_c$, the spectral index $\delta$, and the energy flux $F$. The former two parameters are fixed in each simulation run, while the energy flux rises with time linearly to a peak $F_\mathrm{peak}$ at 10 s, and then falls linearly to zero at 20 s. We have made a grid of 25 simulations by varying $E_c$ from 5 to 25 keV and varying $\delta$ from 3 to 7, where $F_\mathrm{peak}$ is fixed at $10^{10}\ \mathrm{erg}\ \mathrm{s^{-1}}\ \mathrm{cm^{-2}}$. We find that a large value of $E_c$ causes stronger heating at large column depth and a small value of $\delta$ makes the heating more concentrated. The variation of $E_c$ and $\delta$ causes different heating rates with height and different intensities of profiles, while the distribution of line ratios and population ratios do not change qualitatively. Thus, we only show three typical cases here. Besides, we also show another simulation with a larger $F_\mathrm{peak}$ ($10^{11}\ \mathrm{erg}\ \mathrm{s^{-1}}\ \mathrm{cm^{-2}}$) where the line profiles show central reversals. These four cases are described in Table~\ref{four flaring cases}, where the notations are the same as that in \citet{2022Hong}, and the other 22 cases are listed in Appendix~\ref{Simulation lists}. The heating rate of the flaring atmosphere is calculated with the Fokker-Plank approach. We run each simulation for 20 s.

   \begin{table}[htbp]
   \begin{threeparttable}
      \caption{Parameters of flare models\tnote{*}}
      \label{four flaring cases}
\begin{tabular}{|c|c|c|c|}
            \hline
            Case  &  $F_\mathrm{peak}/[\mathrm{erg}\
            \mathrm{s^{-1}}\  \mathrm{cm^{-2}}$] & $E_c/[\mathrm{keV]}$ & $\delta$ \\
            \hline
            $\mathrm{f10E15d3}$ & $10^{10}$  & $15$ & $3$\\
            $\mathrm{f10E25d3}$ & $10^{10}$  & $25$ & $3$ \\
            $\mathrm{f10E25d7}$ & $10^{10}$  & $25$ & $7$ \\
            $\mathrm{f11E25d3}$ & $10^{11}$  & $25$ & $3$ \\
            \hline
         \end{tabular}    
     \begin{tablenotes}
     \footnotesize
        \item[*]{Details for other simulation runs are included in Appendix~\ref{Simulation lists}.}
    \end{tablenotes}
    \end{threeparttable}
   \end{table}

The minority species version of the code (\verb"MS_RADYN") is employed to calculate the \ion{Si}{IV} lines which do not contribute significantly to the atmosphere. We follow the atmosphere of the main run, and solve the level populations and radiative transfer equations for the silicon atom only \citep{2019ApJ...871...23K}. The Si model atom is the same as in \cite{2019ApJ...871...23K}. The snapshot of the simulations are saved every 0.1 s.

\section{Result}
\label{sec3}
    \subsection{The intensity ratio profile $r(\Delta\lambda)$}
    \label{sec3.1}
\begin{figure*}[h]
\centering
\includegraphics[scale=0.165]{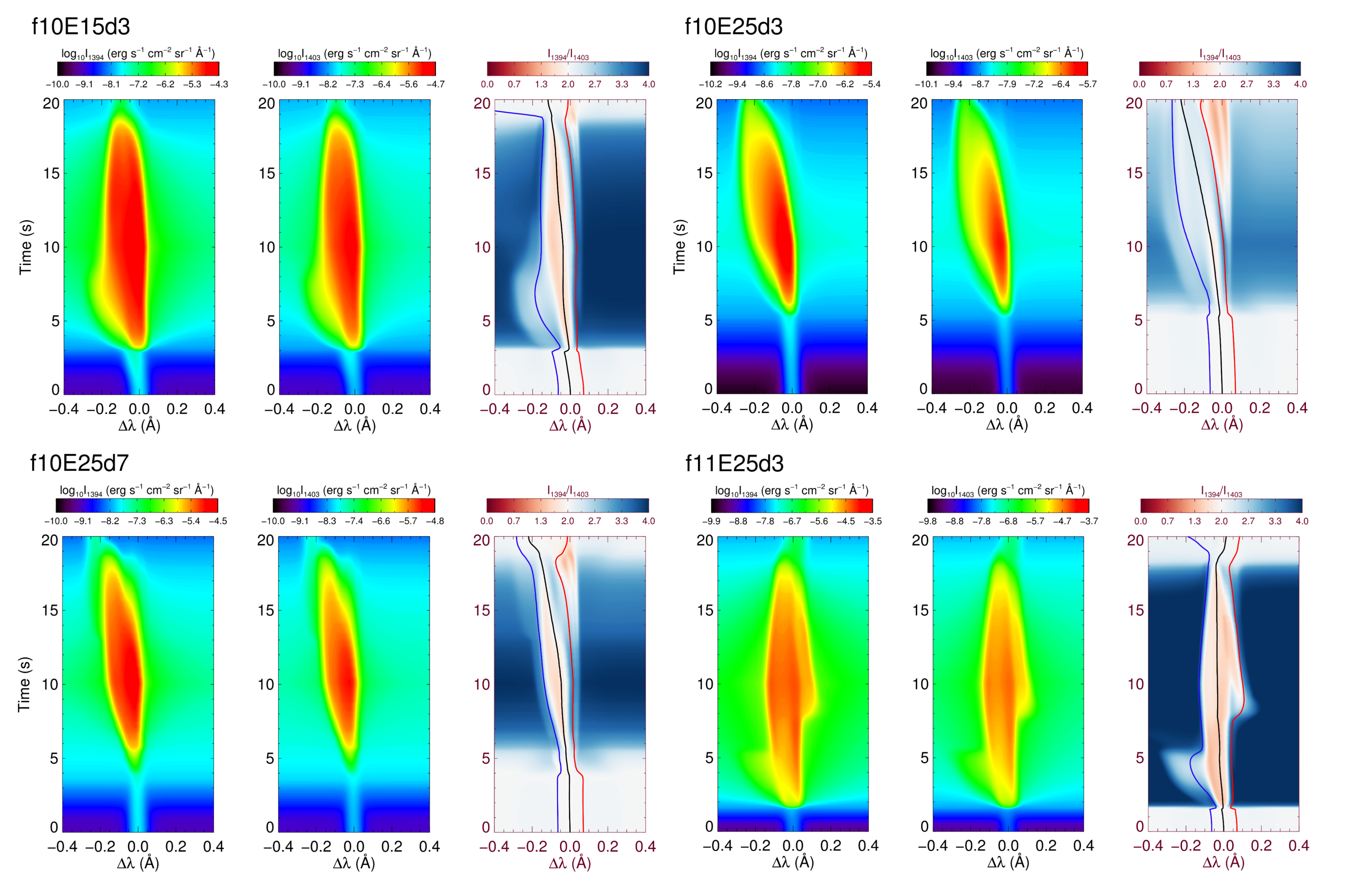}
\caption{Time evolution of the \ion{Si}{IV} line profiles (1394 \AA\ on the left) and the intensity ratio profiles $r(\Delta\lambda)$ in all four flare models. Black, red and blue lines show wavelength positions that are specifically chosen to represent the line core ($\Delta\lambda_c$), the red wing ($\Delta\lambda_r$) and the blue wing ($\Delta\lambda_b$) at each time, respectively, which are defined in Section~\ref{sec3.1}.}
\label{delta}
\end{figure*}

\begin{figure}[h]
\centering
\includegraphics[scale=0.5]{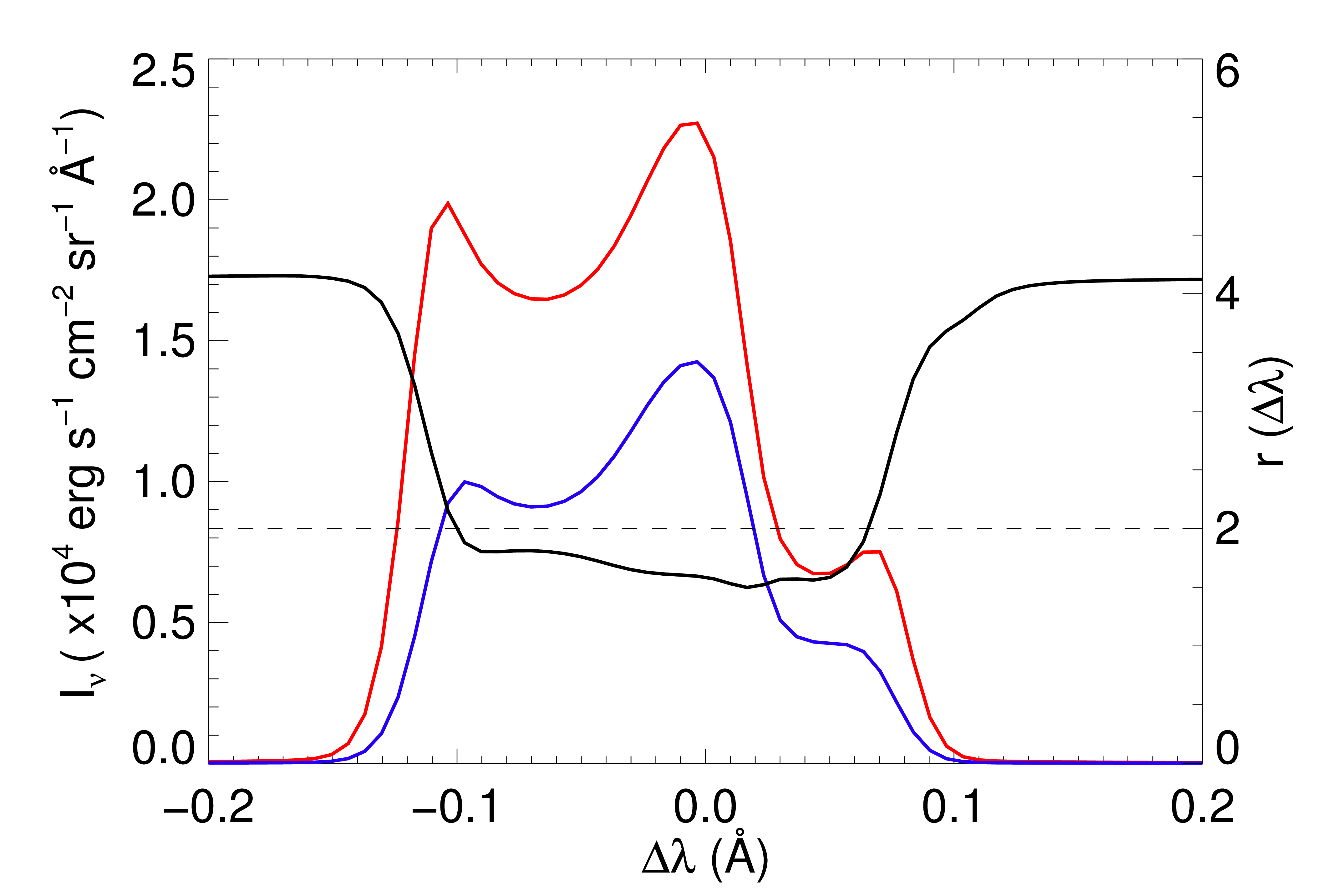}
\caption{Line profiles of \ion{Si}{IV} 1394 \AA\ (red line), \ion{Si}{IV} 1403 \AA\ (blue line) and the intensity ratio profile $r(\Delta\lambda)$ (black line) at $t=10$ s for Case $\mathrm{f11E25d3}$.}
\label{central_reversal}
\end{figure}

The time evolution of the \ion{Si}{IV} intensity profiles and the intensity ratio profile $r(\Delta\lambda)$ for the four simulation cases are shown in Figure~\ref{delta}. At each wavelength, we have subtracted the averaged intensity value at far wings ($\pm 1$ \AA) from the profile to eliminate the influence of the continuum. Such a subtraction is done for every timestep. From the intensity profiles one can see clearly that the \ion{Si}{IV} intensity is enhanced after a certain time of heating, when sufficient energy is deposited in the upper chromosphere ($1.4-1.8$ Mm) and the number of \ion{Si}{IV} atoms is increased. For the first three cases, the line profiles show a single peak that is gradually blueshifted, indicating the upflow of chromospheric evaporation \citep{2019Li,2019ApJ...871...23K}. Comparing the first three panels, we can find some differences in the profiles when changing the spectral index $\delta$ and the cut-off energy $E_c$. A larger value of $\delta$ and a smaller value of $E_c$ mean that the electrons are mostly distributed in the lower-energy part, which cause faster and stronger heating at a low column depth, corresponding to the earlier enhancement and greater intensity, respectively. We note that for Case $\mathrm{f10E25d7}$, the top part of the chromosphere is undisturbed and becomes trapped between the heated chromosphere and the transition region, forming a "chromospheric bubble" \citep{2020Reid}. For Case $\mathrm{f11E25d3}$ (Figure~\ref{central_reversal}), the line profiles show multi-peaks from $t = 5$ s and last for the whole flare. The central reversal caused by absorption in the line core suggests a large opacity at -0.5 \AA, and an additional red component at around 0.8 \AA\ serves as a result of condensation downflows \citep{2022Tian}.

\begin{figure*}[h]
\centering
\includegraphics[scale=0.9]{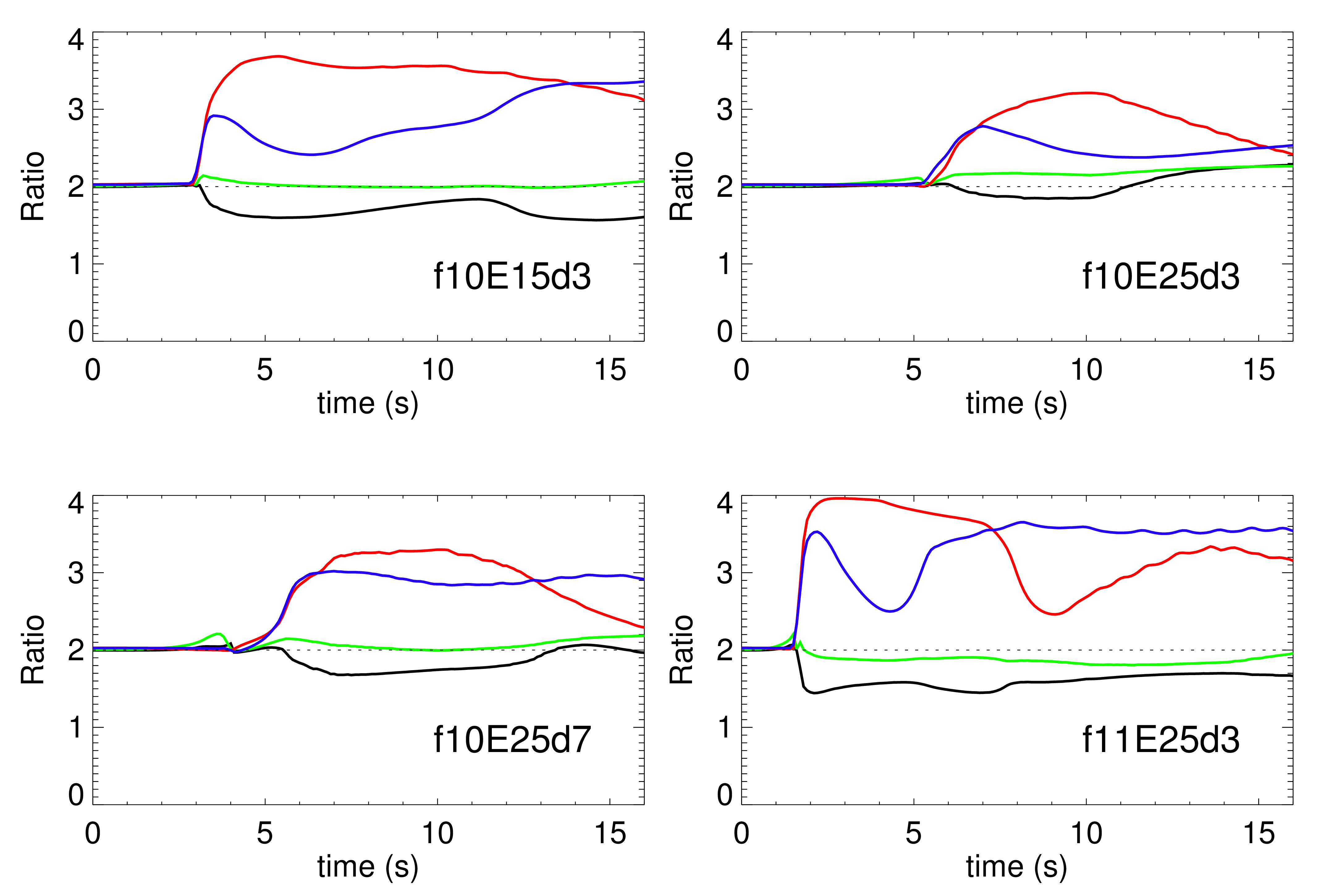}
\caption{Time variations of the ratio at the red wing (red solid line), the blue wing (blue solid line), the line core (black solid line) and the ratio of integrated intensity (green solid line) in four cases. The wavelength positions for the line core, the red wing and the blue wing in each case are marked in Figure~\ref{delta}.}
\label{red_blue}
\end{figure*}

The intensity ratio profile keeps close to 2 before the obvious enhancement of the \ion{Si}{IV} line profiles. A variation of the ratio at different wavelengths is clearly shown, where the ratio at the line wings is generally larger than 2, while the ratio at the line core is generally smaller than 2. For each intensity ratio profile, we choose three specific wavelength positions to represent the line core ($\Delta\lambda_c$), the blue wing ($\Delta\lambda_b$), and the red wing ($\Delta\lambda_r$), respectively, which are overplotted in Figure~\ref{delta}. The centroid of the \ion{Si}{IV} 1403 \AA\ intensity profile is chosen as $\Delta\lambda_c$. The values of  $\Delta\lambda_b$ and $\Delta\lambda_r$ are specifically chosen so that $\int_{\Delta\lambda_\mathrm{min}}^{\Delta\lambda_b}I_{1403\AA}\mathrm{d}\lambda=\int_{\Delta\lambda_{r}}^{\Delta\lambda_\mathrm{max}}I_{1403\AA}\mathrm{d}\lambda=0.01\int_{\Delta\lambda_\mathrm{min}}^{\Delta\lambda_\mathrm{max}}I_{1403\AA}\mathrm{d}\lambda$, where the integration range $[\Delta\lambda_\mathrm{min}, \Delta\lambda_\mathrm{max}]$ is [-0.5 \AA, 0.5 \AA]. In Figure~\ref{red_blue} we show the ratio of wavelength-integrated intensity $R$ and the wavelength-dependent ratio $r(\Delta\lambda)$ at these wavelength positions. It is obvious that after a certain time of heating, $r(\Delta\lambda_b)$ and  $r(\Delta\lambda_r)$ are generally larger than $R$, and $r(\Delta\lambda_c)$ tends to be smaller than $R$. The value of $r(\Delta\lambda)$ can reach as high as 3.5 at the line wings and as low as 1.5 at the line center, while the value of $R$ is still close to 2 owing to an integration effect. The variation behavior of $r(\Delta\lambda)$ and the quantity of $R$ are quite similar to observations at flare ribbons \citep{2022zhou}.

\subsection{The optical thickness}

    Previous studies suggested that the opacity cannot be neglected when we calculate the \ion{Si}{IV} resonance lines during strong flares \citep{2019ApJ...871...23K}. To understand the opacity effect  in line formation, calculation of the contribution function is a useful way \citep{1977A&A....54..227C,1986A&A...163..135M}. The analytic solution of the radiative transfer equation is
    \begin{eqnarray}    
    I_{\lambda,\mu}=\int \frac{1}{\mu}S_{\lambda}(\tau_{\lambda})\mathrm{e}^{-\tau_{\lambda}/\mu} \mathrm{d}\tau_{\lambda}, 
    \end{eqnarray}
    where $\mu=\cos\theta$ and $\theta$ is the angle  between the observer and the vertical direction. We use $\mu\approx1$ in this work. $S_{\lambda}(\tau_{\lambda})$ and $\tau_{\lambda}$ are the source function and the optical depth, respectively. We use the variable $\mathrm{d}z$ (geometrical depth) instead of $\mathrm{d}\tau_{\lambda}$ to get another form: 
    \begin{center}
    \begin{equation}
     I_{\lambda}=\int C_{\lambda} \mathrm{d}z=\int j_{\lambda}(z)\mathrm{e}^{-\tau_{\lambda}(z)}\mathrm{d}z,
     \label{contr}
    \end{equation}
    \end{center}
    where $j_{\lambda}(z)$ is the emissivity, and $C_{\lambda}$ is the contribution function with respect to wavelength per height. We can get the emergent intensity by integrating the contribution function along the light path. 
    
    Hereby we take Case $\mathrm{f10E25d3}$ as an example. The results from other three cases do not show qualitative differences, and are together shown in Appendix~\ref{A}. Figure~\ref{035_contr} shows the contribution function of the \ion{Si}{IV} 1403 \AA\ line. At each wavelength, the contribution function mainly peaks at two heights, contributed by continuum emission from the temperature minimum region and line emission in the upper chromosphere and the transition region, respectively. At $t = 0$ s, the contribution function at the \ion{Si}{IV} 1403 \AA\ line center gathers at $z \approx$ 1.8 Mm, indicating that the line is mainly formed in the transition region. 
    
    The height where $\tau_{\lambda}=1$ is at around 0.5 Mm at this moment, where the continuum is formed, far lower than the transition region. The opacity at the line formation region is quite small where the absorption can be neglected safely, and thus the line can be regarded as optically thin. At $t = 10$ s, the region between 1.5 Mm and 2 Mm that is heated intensively now contributes mostly to the line emission. Meanwhile, the $\tau_{\lambda}=1$ curve also  rises to $z \approx 1.6$ Mm around the line core, which implies that the opacity in the line formation region is now non-negligible \citep{2019ApJ...871...23K}.
    
\begin{figure*}[h]
\centering
\includegraphics[scale=1.0]{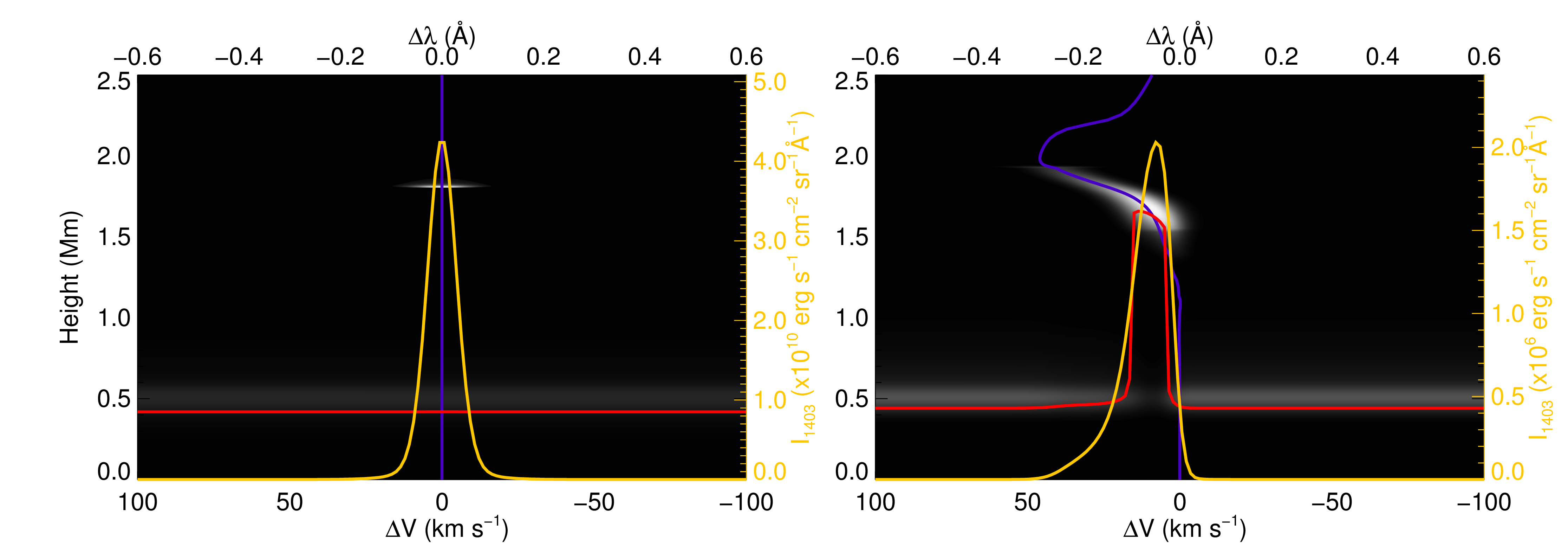}
\caption{Line formation of the \ion{Si}{IV} 1403 \AA\ line in Case $\mathrm{f10E25d3}$, for $t=0$ s on the left panel and $t=10$ s on the right panel. Background gray shades represent the contribution function. Blue lines denote the vertical velocity and red lines denote the $\tau_{\lambda}=1$ height. Orange lines refer to the line profiles.}
\label{035_contr}
\end{figure*}

\begin{figure*}[h]
\centering
\includegraphics[scale=1.0]{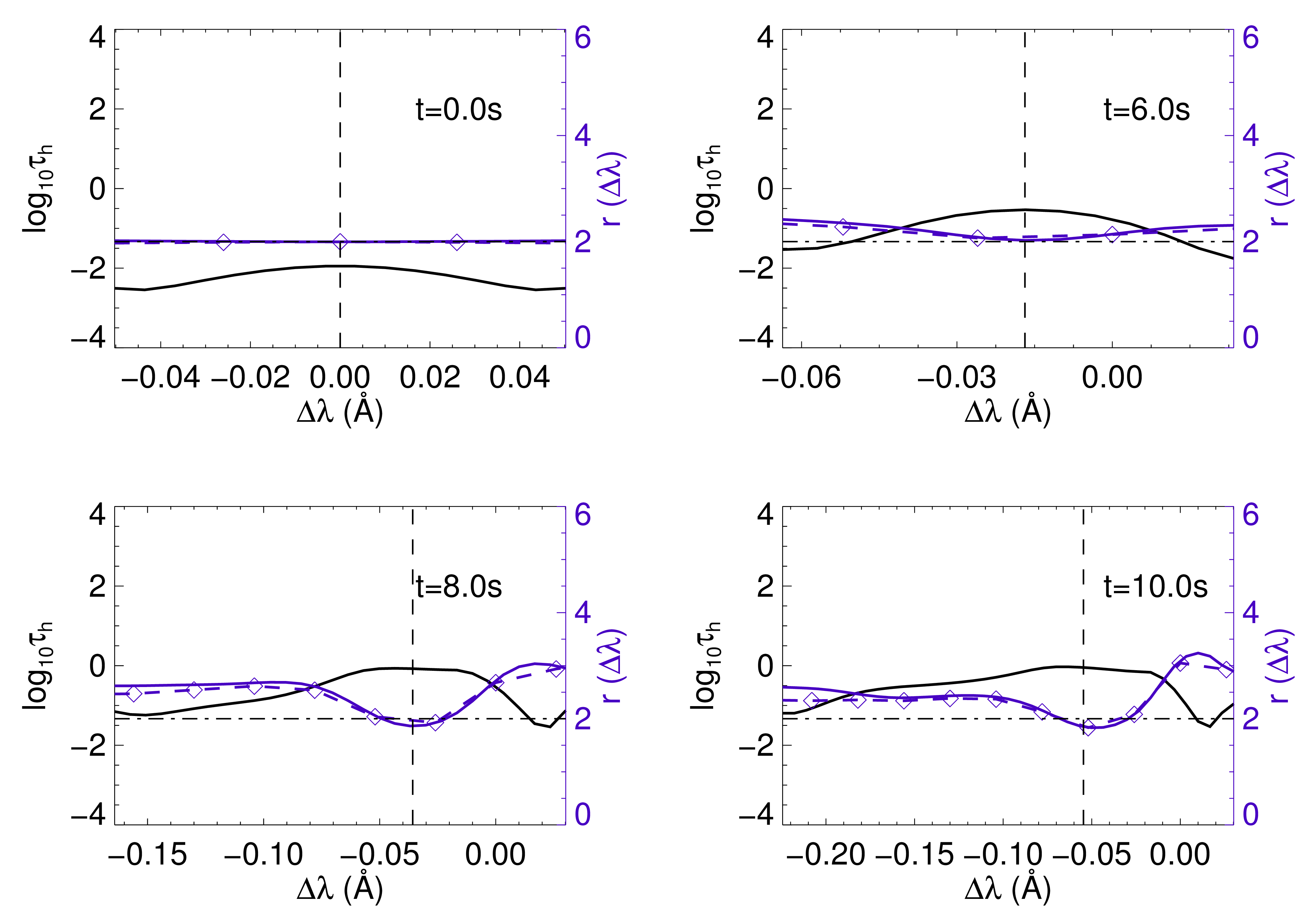}
\caption{The optical depth at the line formation height $\tau_h$ (black line) and the intensity ratio profile $r(\Delta\lambda)$ (blue solid line) for Case $\mathrm{f10E25d3}$. We only show the wavelength range at which the emergent intensity is larger than three times the continuum. Blue dashed line shows $r(\Delta\lambda)$ that are degraded to the IRIS spectral resolution (0.026 \AA). Overplotted are the horizontal line of $r=2$ and the vertical line for the position of the line core.}
\label{ratio_to_tau}
\end{figure*}

We show the intensity ratio profile $r(\Delta\lambda)$ and the optical depth at the line formation height $\tau_h$ as functions of wavelength in Figure~\ref{ratio_to_tau}. The formation height is defined as the centroid of the contribution function \citep{2012Leenaarts}.

    At $t = 0$ s, $\tau_h$ at the line core is smaller than 0.01, which means the opacity effect here is not important, and it is reasonable to treat these lines as optically thin. When the atmosphere is heated during solar flares, $\tau_h$ also increases dramatically, especially at the line core. At $t=10$ s, $\tau_h$ is larger  than 1 at the line core, corresponding to the rise of the $\tau_{\lambda}=1$ curve around the line core in Figure~\ref{delta}. One needs to consider the optical effect at this time although it is still optically thin at the far wings.
    
\begin{figure}[h]
\centering
\includegraphics[scale=0.55]{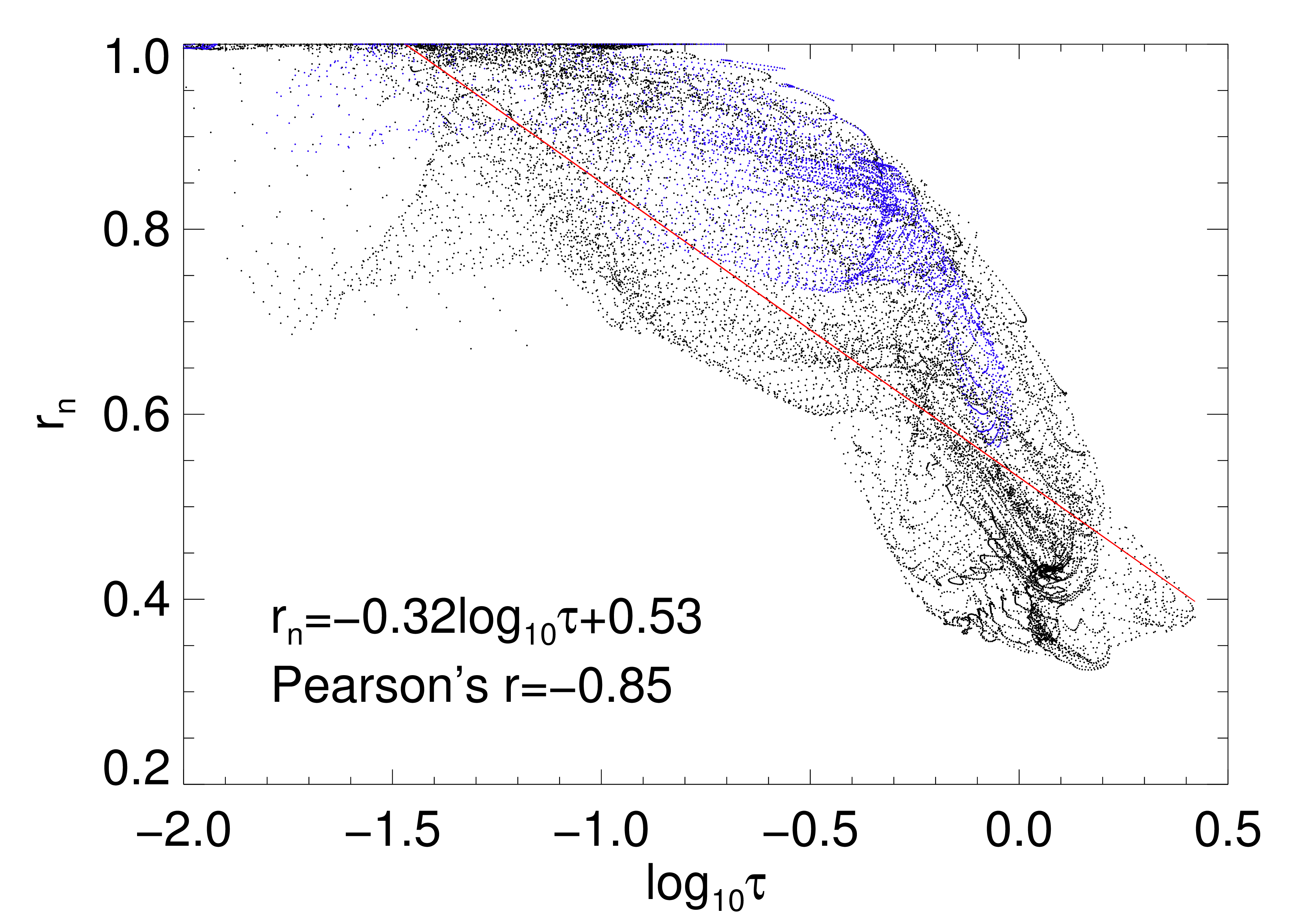}
\caption{Scatter plot of the normalized ratio ($r_n$) versus the optical depth at the formation height ($\tau_h$) for four cases. Blue dots come from Case $\mathrm{f10E25d3}$. The red line shows the result of a linear fit of all the data points.}
\label{combine}
\end{figure}

The variation of $r(\Delta\lambda)$ seems to be negatively correlated to the variation of $\tau_h$. The intensity ratio $r(\Delta\lambda)$ decreases from the line wings to the line core, while the optical depth $\tau_h$ increases. In each intensity ratio profile, the normalized ratio $r_n(\Delta\lambda)$ is defined as: $r_n(\Delta\lambda)=r(\Delta\lambda)/{r_\mathrm{max}}$. In Figure~\ref{combine} we show a scatter plot of the optical depth $\tau_h$ and the normalized ratio $r_n$ at each wavelength point at different times from all four simulation cases. A negative tendency is clearly shown in the figure, which implies that the opacity has an obvious effect on decreasing the intensity ratio. We will discuss the reason for this in Section~\ref{distau}.

\subsection{The population ratio of the \ion{Si}{IV} $3p$ levels}

The two resonance lines of \ion{Si}{IV} share the same lower energy level $3s$ but have different higher levels at $3p$. We hereby label the energy levels at $3s$ ($^{2}$S$_{1/2}$) and $3p$ ($^{2}$P$_{1/2}$ and $^{2}$P$_{3/2}$) as levels 0, 1, and 2, respectively, in the order of increasing energies. Thus, the population densities at the three levels are denoted as $n_0$, $n_1$, and $n_2$, respectively. The two $3p$ levels are so close that ${n_2}/{n_1}$ is considered to be the ratio of their statistical weights in LTE assumption \citep{2015ApJ...811...80R}.

\begin{figure}[h]
\centering
\includegraphics[scale=0.5]{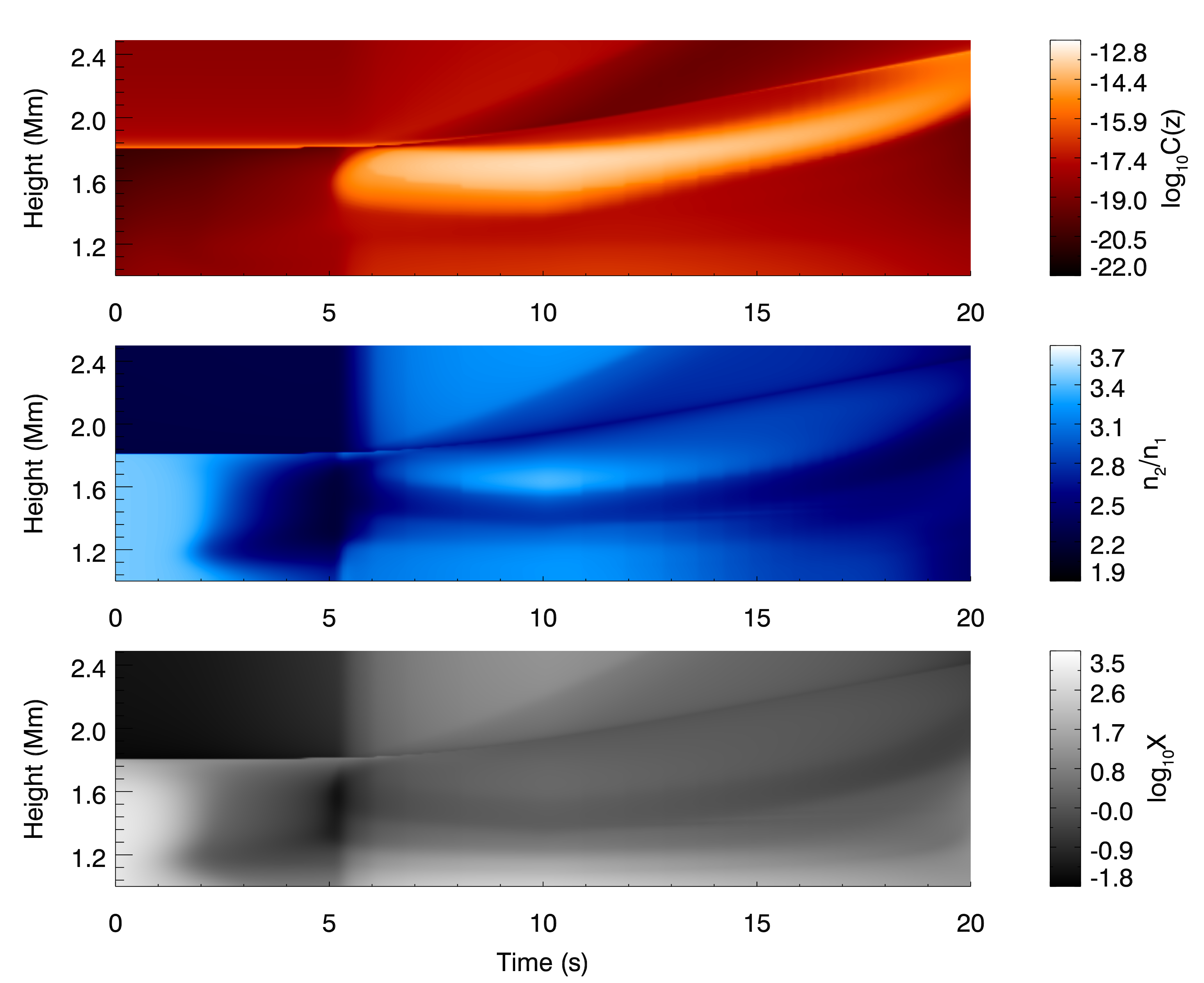}
\caption{Time evolution of height distribution of the contribution function at the line core (top panel), the population ratio $n_2/n_1$ (middle panel) and the proportion of the resonance scattering to thermal emission $X$ (bottom panel) for Case $\mathrm{f10E15d3}$. An obvious enhancement of the contribution function in the chromosphere starts from $t=5.0$ s.}
\label{evolution1}
\end{figure}

We show the population ratio ${n_2}/{n_1}$ in dependence of height and time in Figure~\ref{evolution1}, together with the contribution function at the line core. At $t = 0$ s, the population ratio stays close to 2 in the line formation region (around $1.8-1.9$ Mm). We notice that in the chromosphere the population ratio is much larger than 2 where there is very little contribution to the line intensity. We will discuss why ${n_2}/{n_1}>2$ in Section~\ref{disX}.
       
    As shown in Figure~\ref{delta}, at around $t = 5$ s, the chromosphere is heated to $\sim10^4$ K and the \ion{Si}{IV} line is being enhanced. The contribution function begins to show another peak at around 1.6 Mm from $t = 5.0$ s, and the population ratio above 1.8 Mm starts to increase in the meantime. After $t = 6.0$ s, the contribution function is dominant in the chromosphere, and the population ratio ${n_2}/{n_1}$ is larger than 2 in the layers where the line forms. Checking the results of all four simulation cases, we find an interesting fact that the second $C_{\lambda}$ peak and the deviation of ${n_2}/{n_1}$ from 2 are closely related both spatially and temporxally, which also correspond to the enhancement of \ion{Si}{IV} lines. 
    
\section{Discussion}
\label{sec4}
\subsection{The deviation of the population ratio from 2}
\label{disX}

    The upper levels of the \ion{Si}{IV} 1394 \AA\ and \ion{Si}{IV} 1403 \AA\ lines are the fine structures of \ion{Si}{IV} 3$p$ electron configuration when the degeneracy of the orbital quantum number is broken. The statistical weight is 4 for the $3p$ $^2$P$_{3/2}$ level (the upper level for  the \ion{Si}{IV} 1394 \AA\ transition), and 2 for the $3p$ $^2$P$_{1/2}$ level (the upper level for  the \ion{Si}{IV} 1403 \AA\ transition). Under the LTE assumption, the population ratio ${n_2}/{n_1}$ is described by the Boltzmann equation: 
    \begin{equation}
    \frac{n_2}{n_1}=\frac{g_2}{g_1}\mathrm{e}^{-\frac{\Delta\epsilon}{kT}}.
    \label{boltzmann}
    \end{equation}   
    where ${g_1}$ and ${g_2}$ are statistical weights of level 1 and level 2, respectively. The energy level difference $\Delta\epsilon$ is so small that $\mathrm{e}^{-\frac{\Delta\epsilon}{kT}}\sim1$. Hence, the population ratio ${n_2}/{n_1}$ is considered to be equal to ${g_2}/{g_1}$, which is 2 exactly. However, the LTE assumption is no longer valid above the photosphere. At the chromosphere and the transition region, the non-LTE effect is dominant. 

We write the population equations for levels 1 and 2 (the $3p$ levels) as: 
    \begin{eqnarray}
    \label{state1}
    \frac{\mathrm{d} n_1}{\mathrm{d} t}&=&n_{0}R_{01}+n_{0}C_{01}+n_{2}C_{21}+n_{c}R_{c1}+n_{c}C_{c1}\nonumber\\
  &  &-n_1(R_{10}+C_{10}+C_{12}+R_{1c}+C_{1c}),\\    
    \frac{\mathrm{d} n_2}{\mathrm{d} t}&=&n_{0}R_{02}+n_{0}C_{02}+n_{1}C_{12}+n_{c}R_{c2}+n_{c}C_{c2}\nonumber\\
   & &-n_2(R_{20}+C_{20}+C_{21}+R_{2c}+C_{2c}),
    \label{state2}
    \end{eqnarray}
    where $R_{ij}$ and $C_{ij}$ denote the radiative and collisional transition rates, and the subscript $c$ represents the continuum level of \ion{Si}{IV}.

In Figure~\ref{compare} we compare all the transition rates in Equation~\ref{state1} at 6.0 s, where $n_1R_{10}=n_1A_{10}+n_1B_{10}\Bar{J}_{10}$. It is very clear that in the line formation region, the transition rates between level 0 and level 1 are orders of magnitude larger than the rates of other transitions. The dominant terms are collisional and radiative excitations plus spontaneous emission. Considering that the time derivatives of the level populations are very small (see Figure~\ref{compare}), we further simplify Equations \ref{state1} and \ref{state2} by including only the dominant terms:

\begin{figure}[h]
\centering
\includegraphics[scale=0.5]{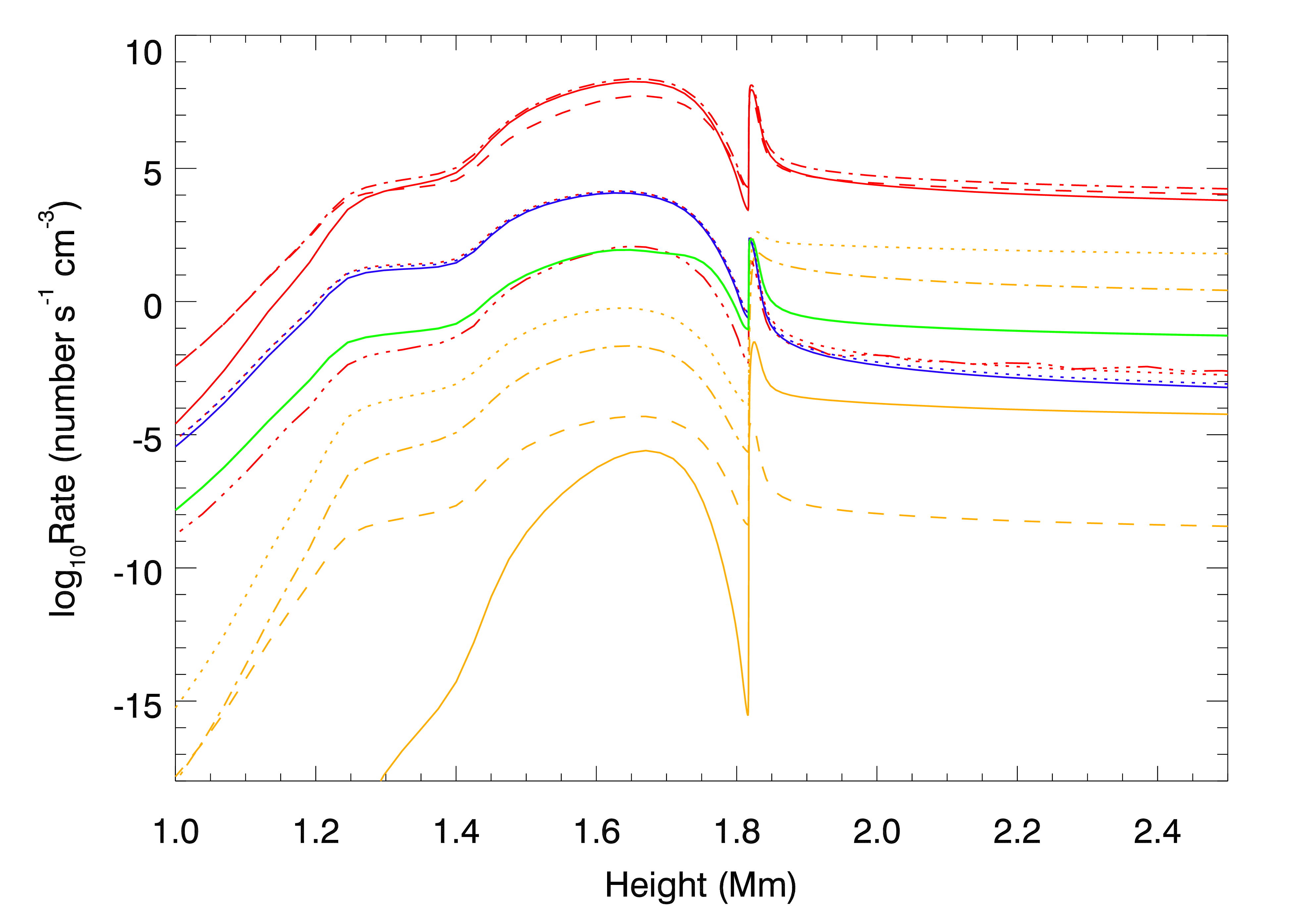}
\caption{Height distribution of all the transition rates in Equation~\ref{state1} for Case $\mathrm{f10E25d3}$ at 6.0 s. Red lines represent transitions between levels 0 and 1: $n_0C_{01}$ (red solid line), $n_1C_{10}$ (red dotted line), $n_0R_{01}$ (red dashed line), $n_1A_{10}$ (red dotted dashed line), $n_1B_{10}\Bar{J}_{10}$ (red three-dotted dashed line). Blue lines represent transitions between levels 1 and 2: $n_1C_{12}$ (blue solid line), $n_2C_{21}$ (blue dotted line). Yellow lines represent transitions between level 1 and the continuum: $n_1C_{1c}$ (yellow solid line), $n_cC_{c1}$ (yellow dotted line), $n_1R_{1c}$ (yellow dashed line), $n_cR_{c1}$ (yellow dotted dashed line). The net rate $\mathrm{d}n_1/\mathrm{d}t$ is plotted as the green line.}

\label{compare}
\end{figure}
    
    \begin{eqnarray}   
    n_0B_{01}\Bar{J}_{01}+n_0C_{01}-n_1A_{10}=0,\\
    \label{state3}
    n_0B_{02}\Bar{J}_{02}+n_0C_{02}-n_2A_{20}=0.
    \label{state4}
    \end{eqnarray}
    The transitions can be divided into two processes: collisional excitation followed by radiative deexcitation, which corresponds to thermal emission; radiative excitation followed by radiative deexcitation, which is referred to as resonance scattering \citep{2018A&A...619A..64G}. Assuming that $\Bar{J}_{02}=(1+k)\Bar{J}_{01}\;(k>0)$ and noticing that $A_{20}/A_{10}=1$, $C_{02}/C_{01}=B_{02}/B_{01}=2$, we obtain
    
    \begin{equation}
    \frac{n_2}{n_1}=2+\frac{2kX}{1+X},
    \label{state5}
    \end{equation}
    where $X=B_{01}\bar{J}_{01}/C_{01}$ is the proportion of the resonance scattering to thermal emission. We can find that the population ratio is positively correlated with the value of $X$, that is, the increase of resonance scattering makes $n_2/n_1>2$.


    The above mechanism is illustrated in the bottom panel of Figure~\ref{evolution1}. At $t = 0$ s, the radiation is small enough ($X\ll1$) in the line formation region ($1.8-1.9$ Mm), and the population ratio is 2 according to Equation~\ref{state5}, which is also the result under coronal approximation. After a certain time of flare heating (5 s in Case $\mathrm{f10E25d3}$), the local radiation is enhanced and the radiative excitation rate is comparable to the collisional excitation rate. The population ratio also increases gradually with time and becomes larger than 2 in the line formation region (Figure~\ref{evolution1}).

\subsection{The deviation of the intensity ratio from 2}
\label{distau}
We now discuss how would the population ratio influences the line intensity ratio. We first consider the optically thin case  where the optical depth is so small that $\mathrm{e}^{-\tau_{\lambda}}\approx1$, and $C_\lambda\approx j_\lambda$. Thus we can obtain the emergent intensity by only integrating the emissivity along the height:

    \begin{equation}
    I_{1403\AA}(\Delta\lambda)\approx\int j_{1403\AA}(z,\Delta\lambda)\mathrm{d}z=\int n_{1}(z)A_{10}(z)\psi_{1403\AA}(z,\Delta\lambda)\mathrm{d}z,
    \label{thin_1}
    \end{equation}
    \begin{equation}
    I_{1394\AA}(\Delta\lambda)\approx\int j_{1394\AA}(z,\Delta\lambda)\mathrm{d}z=\int n_{2}(z)A_{20}(z)\psi_{1394\AA}(z,\Delta\lambda)\mathrm{d}z,
    \label{thin_2}
    \end{equation}
where $\psi(z,\Delta\lambda)$ are normalized profiles. Noticing that $A_{20}=A_{10}$, $\psi_{1394\AA}(z,\Delta\lambda)\approx\psi_{1403\AA}(z,\Delta\lambda)$, the ratio of emergent intensity only depends on the population ratio ${n_2}/{n_1}$. 
At $t=0$ s, the \ion{Si}{IV} lines are optically thin, and the intensity ratio stays close to 2 (Figure~\ref{ratio_to_tau}), which is equal to the population ratio ${n_2}/{n_1}$ (Figure~\ref{evolution1}). 

    In flare conditions, we need to consider the optical depth term in the contribution function. The intensity ratio is related to the ratio of the contribution functions since the integration range is the same. The ratio of contribution functions is given as
    \begin{equation}   
    \frac{C_{1394\AA}}{C_{1403\AA}}=\frac{j_{1394\AA}\mathrm{e}^{-\tau_{1394\AA}}}{j_{1403\AA}\mathrm{e}^{-\tau_{1403\AA}}}=\frac{n_2}{n_1}\mathrm{e}^{-(\tau_{1394\AA}-\tau_{1403\AA})},
    \label{C_ratio}
    \end{equation}
which is a function of wavelength and height. As we has shown above, deviation of the ratio $n_2/n_1$ from 2 is mainly from the resonance scattering effect, while the term $\mathrm{e}^{-(\tau_{1394\AA}-\tau_{1403\AA})}$ apparently represents the opacity effect. At each height, the ratio of contribution function is reduced by the opacity effect because $\tau_{1394\AA}>\tau_{1403\AA}$. As a result, the intensity ratio is reduced by the opacity effect. However, when the population ratio ${n_2}/{n_1}$ grows due to resonance scattering, the intensity ratio increases proportionally. These two effects compete with each other. Besides their opposite effects on the magnitude of the intensity ratio, the effect of ${n_2}/{n_1}$ is wavelength independent while that of $\tau$ is wavelength dependent. 

The intensity ratio profiles in Figure~\ref{delta} could then be explained as follows. The resonance scattering effect elevates the intensity ratio $r(\Delta\lambda)$ to larger than 2, while the opacity effect pulls the ratio back to some extent. At the line center, the opacity is relatively larger so that the intensity ratio is pulled back by a larger amount, causing a U-shaped profile. The ratio at the line wings are generally larger than 2, while the ratio at the line core could be larger or smaller than 2, depending on how much effect the opacity exerts on the ratio.
    
We further illustrate these two effects by defining the difference of contribution functions as $\Delta C(\Delta\lambda,z)=C_{1394\AA}(\Delta\lambda,z)-2C_{1403\AA}(\Delta\lambda,z)$. The integration of $\Delta C$ along height gives the difference of intensities $I_{1394\AA}(\Delta\lambda)-2 I_{1403\AA}(\Delta\lambda)$. Therefore, the value of $\Delta C$ indicates the contribution to the deviation of $r(\Delta\lambda)$ from 2. A positive $\Delta C$ tends to elevate $r(\Delta\lambda)$ to over 2, while a negative one tends to suppress $r(\Delta\lambda)$ to below 2.

\begin{figure}[h]
\centering
\includegraphics[scale=0.45]{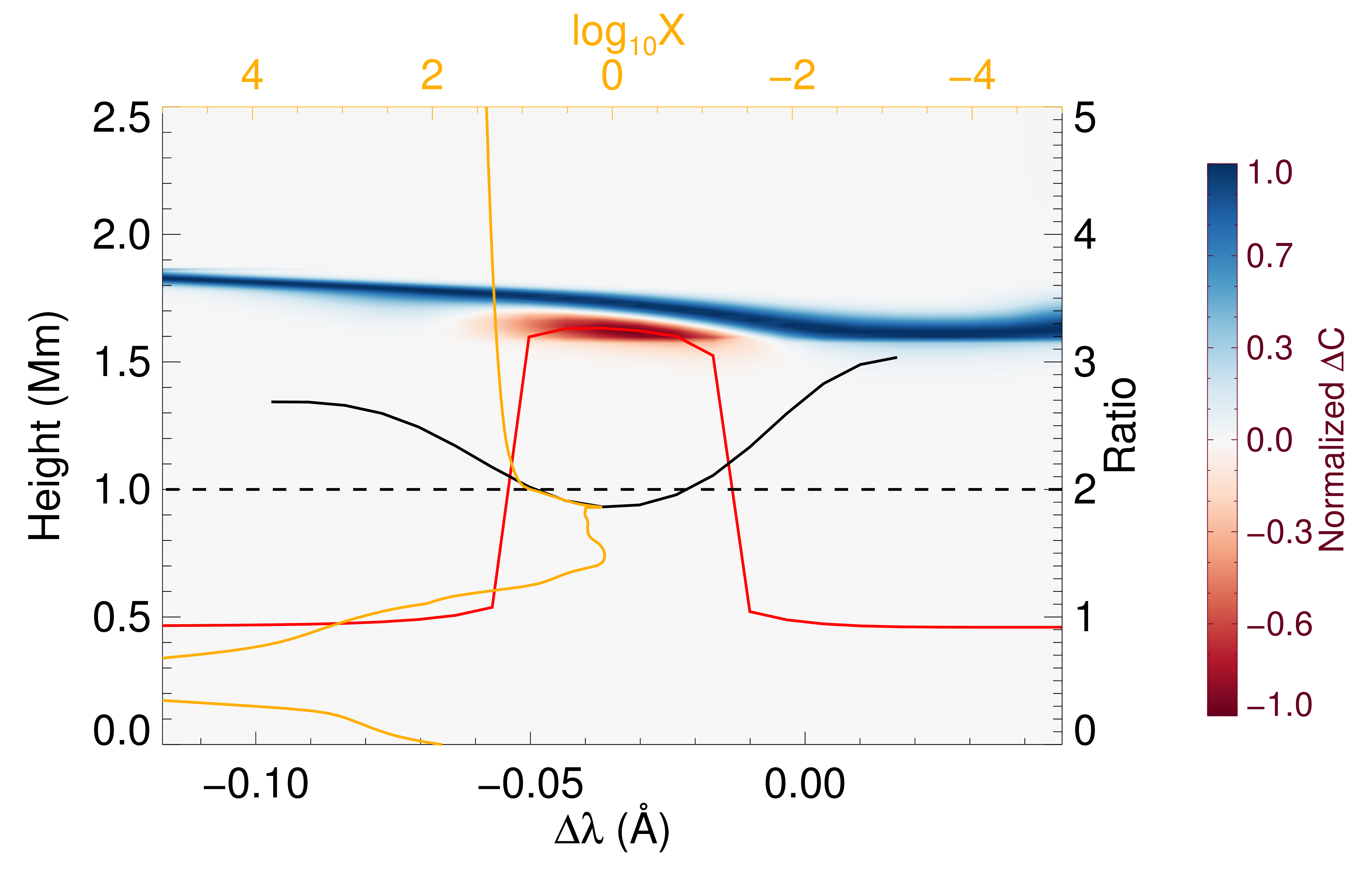}
\caption{Normalized difference of contribution functions ($\Delta C$) at 8.0 s in Case $\mathrm{f10E25d3}$ (background). The intensity ratio profile is marked as the black line. The red line denotes the $\tau =1$ height and the orange line refers to the proportion of resonance scattering to thermal emission ($X$).}
\label{035_RC}
\end{figure}

In Figure~\ref{035_RC} we show the distribution of $\Delta C$ at $t =8.0$ s for Case $\mathrm{f10E25d3}$. Note that $\Delta C$ is normalized at each wavelength. At each wavelength, the region where $\Delta C>0$ is located in the upper chromosphere (1.6--1.9 Mm). It is also the line formation region where the contribution function $C_\lambda$ is large and resonance scattering is strong ($X>1$). However, the region where $\Delta C<0$ only appears in the line core, coinciding with the region with a large opacity.

    Thus, the $r(\Delta\lambda)$ profile shows a U-shape, in which the ratio increases towards the line wings but decreases towards the line core. Similar features of $r(\Delta\lambda)$ and $\Delta C$ are found in other simulation cases during flare heating.

\subsection{Comparison with observations}

    In Figure~\ref{ratio_to_tau}, we also plot the ratio of the intensity profiles that are degraded to the IRIS spectral resolution (0.026 \AA). One can see that the U-shaped profile is still obvious, with the value at the line core smaller than 2, while the value at the line wings larger than 2. Actually, such U-shaped profiles have been reported in previous observations \citep{2022zhou}.
    
    The central reversal in emission lines is often caused by the large line opacity. In observations there are reports of such features in the \ion{Si}{IV} lines, especially in flares \citep{2015ApJ...811...48Y, 2022zhou,2022Juraj}. In our simulations, we find centrally reversed profiles in Case $\mathrm{f11E25d3}$ (Figure~\ref{central_reversal}), where the energy flux $F$ is one order of magnitude higher than that in other three cases. We confirm that the central reversal is due to the increased line opacity.
    
In many observations of the Sun or other stars, the ratio of integrated intensity $R$ is taken as an indicator of the optical thickness \citep{1999A&A...351L..23M, 2006Christian, 2015ApJ...811...48Y, 2015Brannon,2021Sargam}. \cite{2022zhou} claimed that the wavelength-dependent ratio profile $r(\Delta\lambda)$ would be a better indicator since the opacities at the line core and at the line wings are different. Our results confirm that the intensity ratio profile $r(\Delta\lambda)$ usually has a U-shape as in observations. However, we find it still insufficient in some cases to use the value of either $R$ or $r(\Delta\lambda)$ as an indicator of the optical thickness. For example, in the simulation snapshot of Case $\mathrm{f10E25d3}$ at 6.0 s, both $R$ and the intensity ratio at the line core $r(\Delta\lambda_c)$ are larger than 2, but we do see an obvious increase in the line opacity (Figure~\ref{ratio_to_tau}). The normalized ratio $r_n$ might be a complementary indicator of the optical thickness as judged from Figure~\ref{combine}. Quantitatively, the \ion{Si}{IV} line is not optically thin any more if $r_n<0.6$.


\section{Conclusions}
\label{sec5}
In this paper, we analyze the properties of the two \ion{Si}{IV} resonance lines in flare conditions based on \verb"RADYN" simulations. We focus on the intensity ratio profile $r(\Delta\lambda)$ and find that the value of $r(\Delta\lambda)$ rises at the line wings and falls at the line core. By comparison, the ratio of integrated intensity $R$ lies between the minimum and maximum of $r(\Delta\lambda)$. At most of the wavelength points, $r(\Delta\lambda)$ deviates from 2 obviously, ranging from 1.5 to 4. However, $R$ usually ranges from 1.8 to 2.3. We agree with \cite{2019ApJ...871...23K} that the opacity of the two resonance lines are both non-negligible when flare occurs. In the line core, the line opacity is larger and  $r(\Delta\lambda)$ becomes smaller than that in the line wings. By calculating the normalized ratio $r_n(\Delta\lambda)=r(\Delta\lambda)/{r_\mathrm{max}}$, we can quantitatively estimate the optical depth at the line formation height through an empirical relationship between $r_{n}$ and $\tau$ (Figure \ref{combine}). 

We also find that due to flare heating, the increased rate of resonance scattering will increase the population ratio of the \ion{Si}{IV} $3p$ levels ($n_2/n_1$), which will in turn increase the intensity ratio $r(\Delta\lambda)$. In the mean time,  the large opacity at the line core tends to decrease the intensity ratio $r(\Delta\lambda)$. The competition of the resonance scattering effect and the opacity effect will result in a U-shaped intensity ratio profile as in observations.

As noted above, the values of $R$ and $r(\Delta\lambda)$ are influenced by both resonance scattering and opacity. In the case that $R$ and $r(\Delta\lambda)$ are not sufficient to judge the optical thickness, we propose to use the normalized ratio $r_n(\Delta\lambda)$ as a complementary criterion. We conclude that if $r_n<0.6$, the line should be regarded as optically thick. Such a case most likely appears at the line core. On the other hand, since the opacity at the far wings is always small, deviation of the intensity ratio $r(\Delta\lambda)$ from 2 can safely serve as an indicator of the strength of resonance scattering.

%
%

\begin{acknowledgements}
We are grateful to the referee for careful reading of the paper and constructive comments. We would like to thank Graham Kerr for providing the \ion{Si}{IV} model atom. This work was supported by National Key R\&D Program of China under grant 2021YFA1600504 and by NSFC under grants 11903020 and 12127901.

\end{acknowledgements}

\bibliography{citation}{}
\bibliographystyle{aa}

\begin{appendix}
 
\section{Additional Simulation Cases}

\renewcommand{\thetable}{A.\arabic{table}} 
\setcounter{table}{0}
     \label{Simulation lists}
    \begin{table}[htbp]
    \caption{Additional 22 simulation cases}

    \begin{tabular}{|c|c|c|c|}        
            \hline
            Case  &  $F_\mathrm{peak}/[\mathrm{erg}\
            \mathrm{s^{-1}}\  \mathrm{cm^{-2}}$] & $E_c/[\mathrm{keV]}$ & $\delta$ \\
            \hline

            $\mathrm{f10E05d3}$ & & $05$ & $3$ \\
            $\mathrm{f10E10d3}$ & & $10$ & $3$ \\            

            $\mathrm{f10E20d3}$ & & $20$ & $3$ \\

            $\mathrm{f10E05d4}$ & & $05$ & $4$ \\
            $\mathrm{f10E10d4}$ & & $10$ & $4$ \\            
            $\mathrm{f10E15d4}$ & & $15$ & $4$ \\
            $\mathrm{f10E20d4}$ & & $20$ & $4$ \\
            $\mathrm{f10E25d4}$ & & $25$ & $4$ \\            
            $\mathrm{f10E05d5}$ & & $05$ & $5$ \\
            $\mathrm{f10E10d5}$ & & $10$ & $5$ \\            
            $\mathrm{f10E15d5}$ & $10^{10}$  & $15$ & $5$ \\
            $\mathrm{f10E20d5}$ & & $20$ & $5$ \\
            $\mathrm{f10E25d5}$ & & $25$ & $5$ \\
            $\mathrm{f10E05d6}$ & & $05$ & $6$ \\
            $\mathrm{f10E10d6}$ & & $10$ & $6$ \\            
            $\mathrm{f10E15d6}$ & & $15$ & $6$ \\
            $\mathrm{f10E20d6}$ & & $20$ & $6$ \\
            $\mathrm{f10E25d6}$ & & $25$ & $6$ \\
            $\mathrm{f10E05d7}$ & & $05$ & $7$ \\
            $\mathrm{f10E10d7}$ & & $10$ & $7$ \\            
            $\mathrm{f10E15d7}$ & & $15$ & $7$ \\
            $\mathrm{f10E20d7}$ & & $20$ & $7$ \\
 
            \hline
         \end{tabular}
    \end{table}


    \section{Additional Figures}

\renewcommand{\thefigure}{B.\arabic{figure}}
\setcounter{figure}{0}
     \label{A}
    \begin{figure}[h]
    \centering
    \includegraphics[scale=0.46]{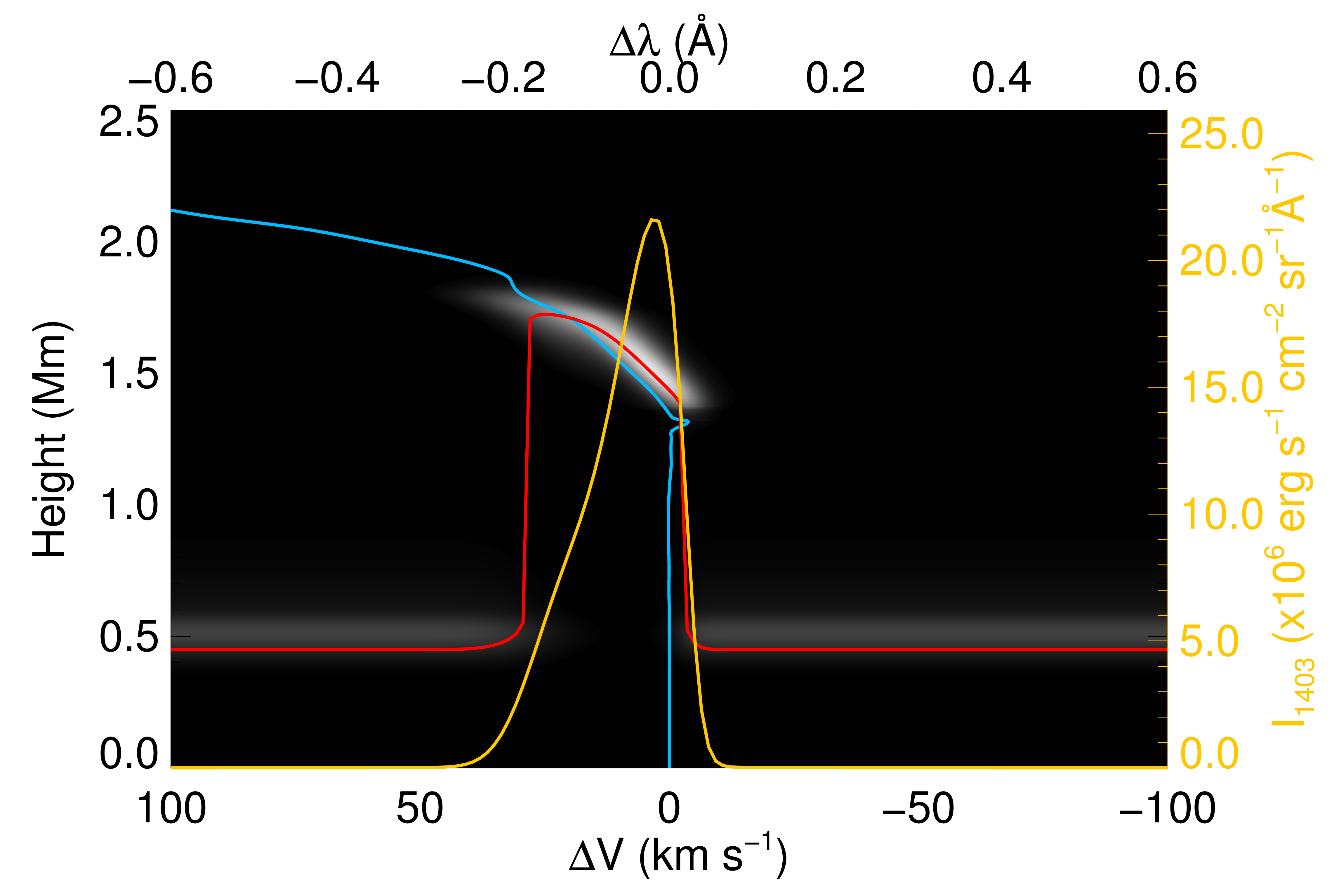}
    \caption{The same as Figure~\ref{035_contr} but for Case $\mathrm{f10E15d3}$ at $t=10$ s.}
    \end{figure}  

    \begin{figure}[h]
    \centering
    \includegraphics[scale=0.46]{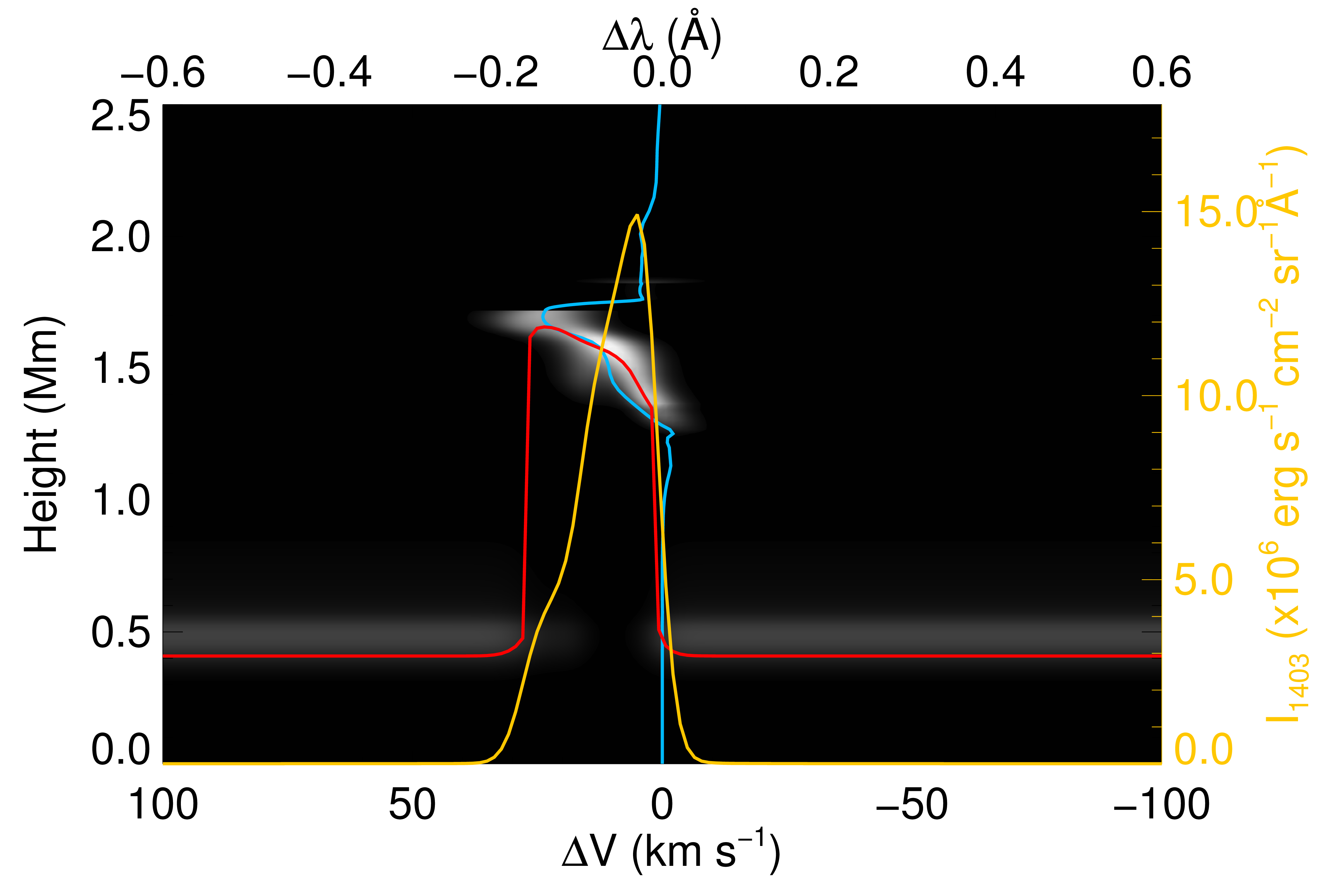}
    \caption{The same as Figure~\ref{035_contr} but for Case $\mathrm{f10E25d7}$ at $t=10$ s.}
    \end{figure} 

    \begin{figure}[h]
    \centering
    \includegraphics[scale=0.46]{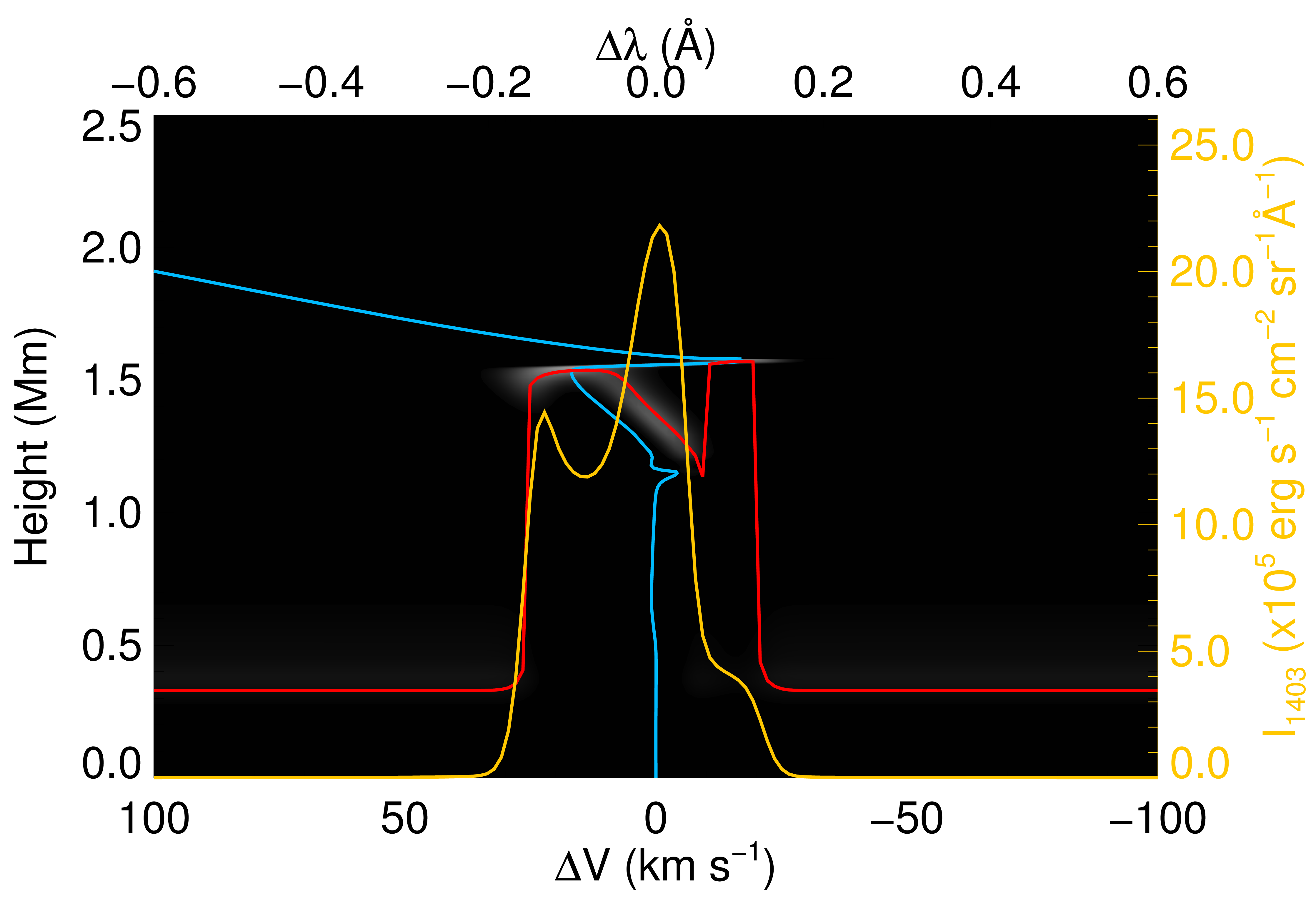}
    \caption{The same as Figure~\ref{035_contr} but for Case $\mathrm{f11E25d3}$ at $t=10$ s.}
    \end{figure} 
    
    \begin{figure}[h]
    \centering
    \includegraphics[scale=0.5]{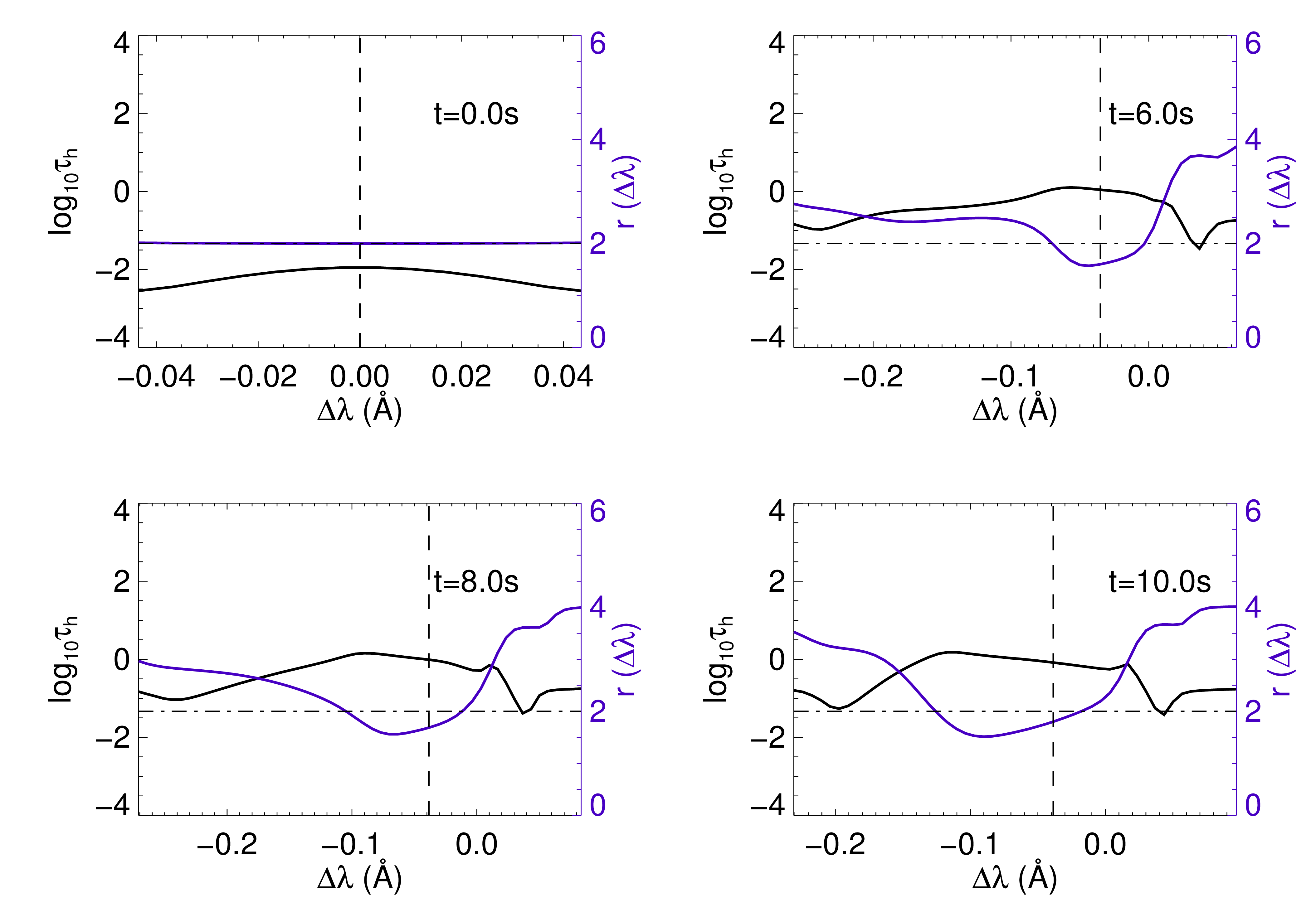}
    \caption{The same as Figure~\ref{ratio_to_tau} but for Case $\mathrm{f10E15d3}$.}
    \end{figure} 

    \begin{figure}[h]
    \centering
    \includegraphics[scale=0.5]{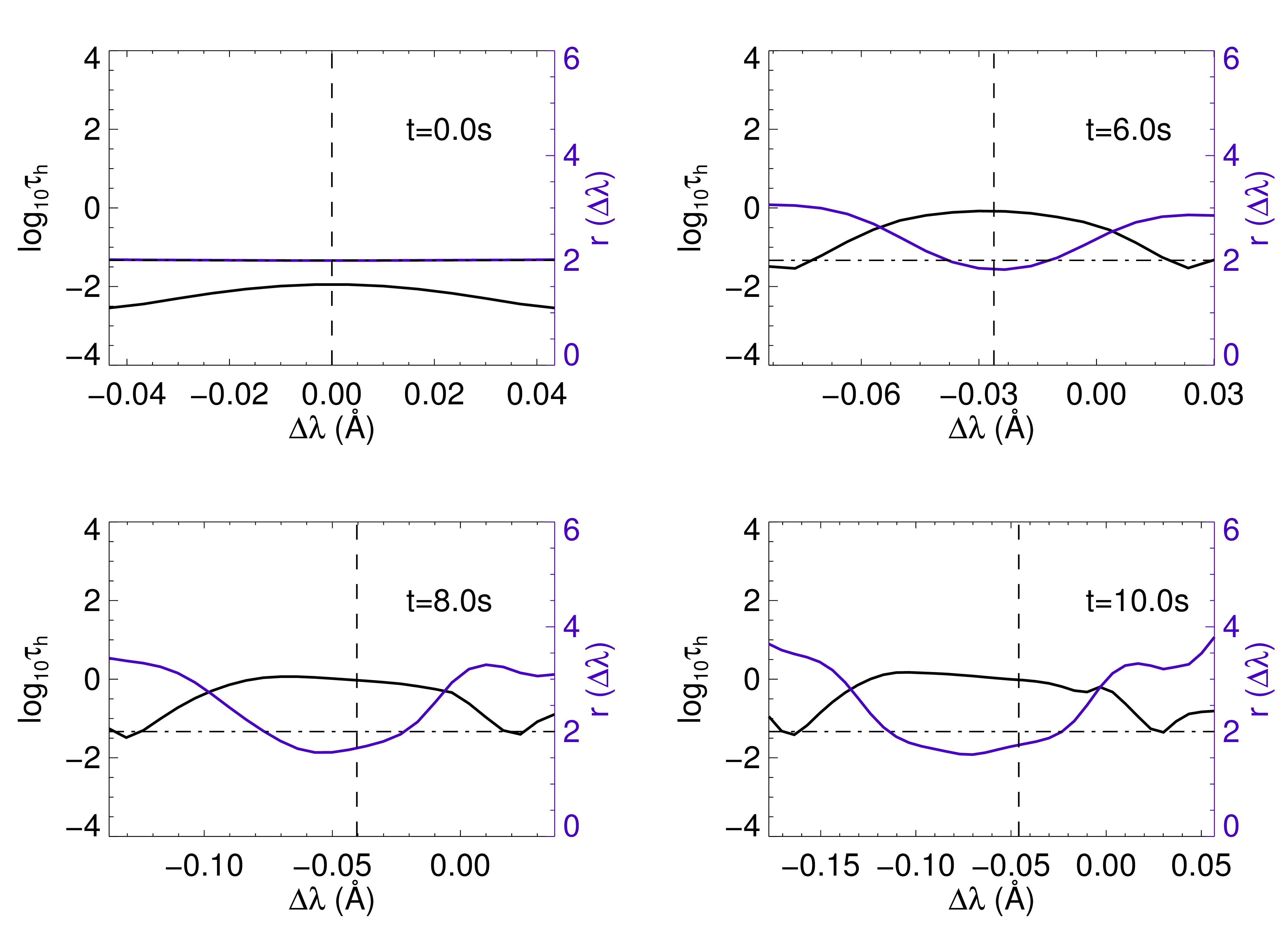}
    \caption{The same as Figure~\ref{ratio_to_tau} but for Case $\mathrm{f10E25d7}$.}
    \end{figure}  

    \begin{figure}[h]
    \centering
    \includegraphics[scale=0.5]{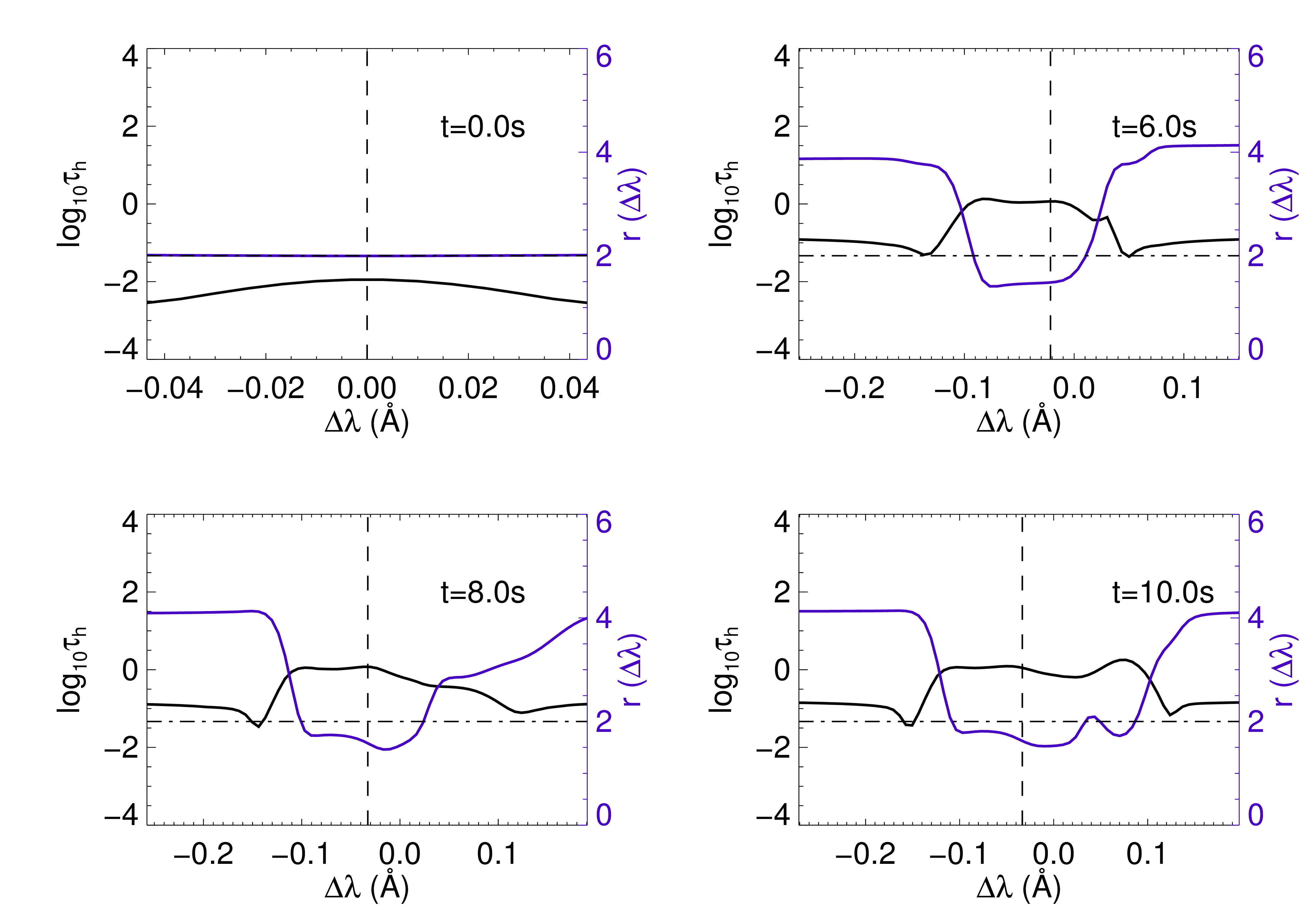}
    \caption{The same as Figure~\ref{ratio_to_tau} but for Case $\mathrm{f11E25d3}$.}
    \end{figure} 

    \begin{figure}[h]
    \centering
    \includegraphics[scale=0.48]{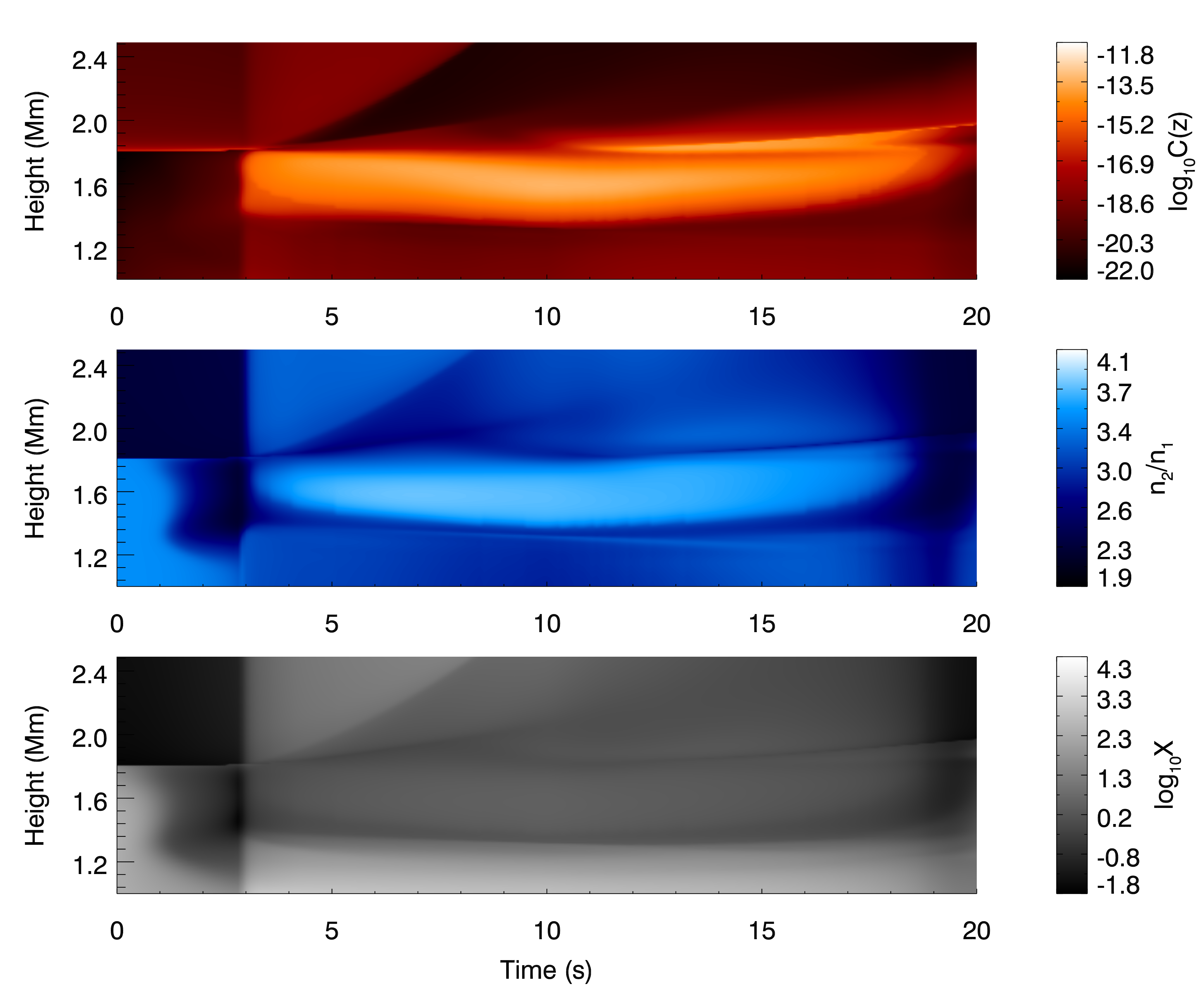}
    \caption{The same as Figure~\ref{evolution1} but for Case $\mathrm{f10E15d3}$.}
    \label{033_evolution}
    \end{figure} 

    \begin{figure}[h]
    \centering
    \includegraphics[scale=0.48]{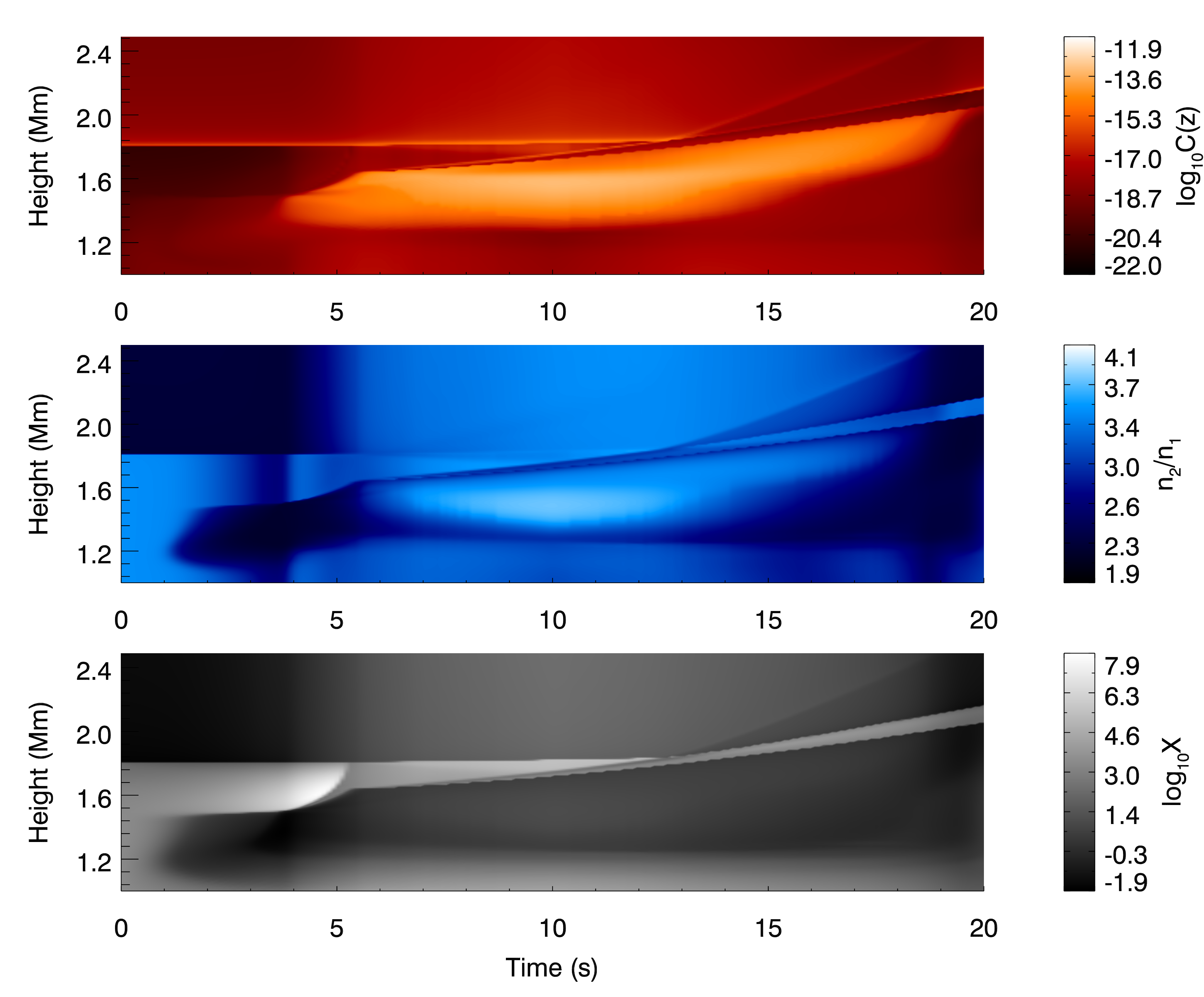}
    \caption{The same as Figure~\ref{evolution1} but for Case $\mathrm{f10E25d7}$.}
    \end{figure} 

    \begin{figure}[h]
    \centering
    \includegraphics[scale=0.48]{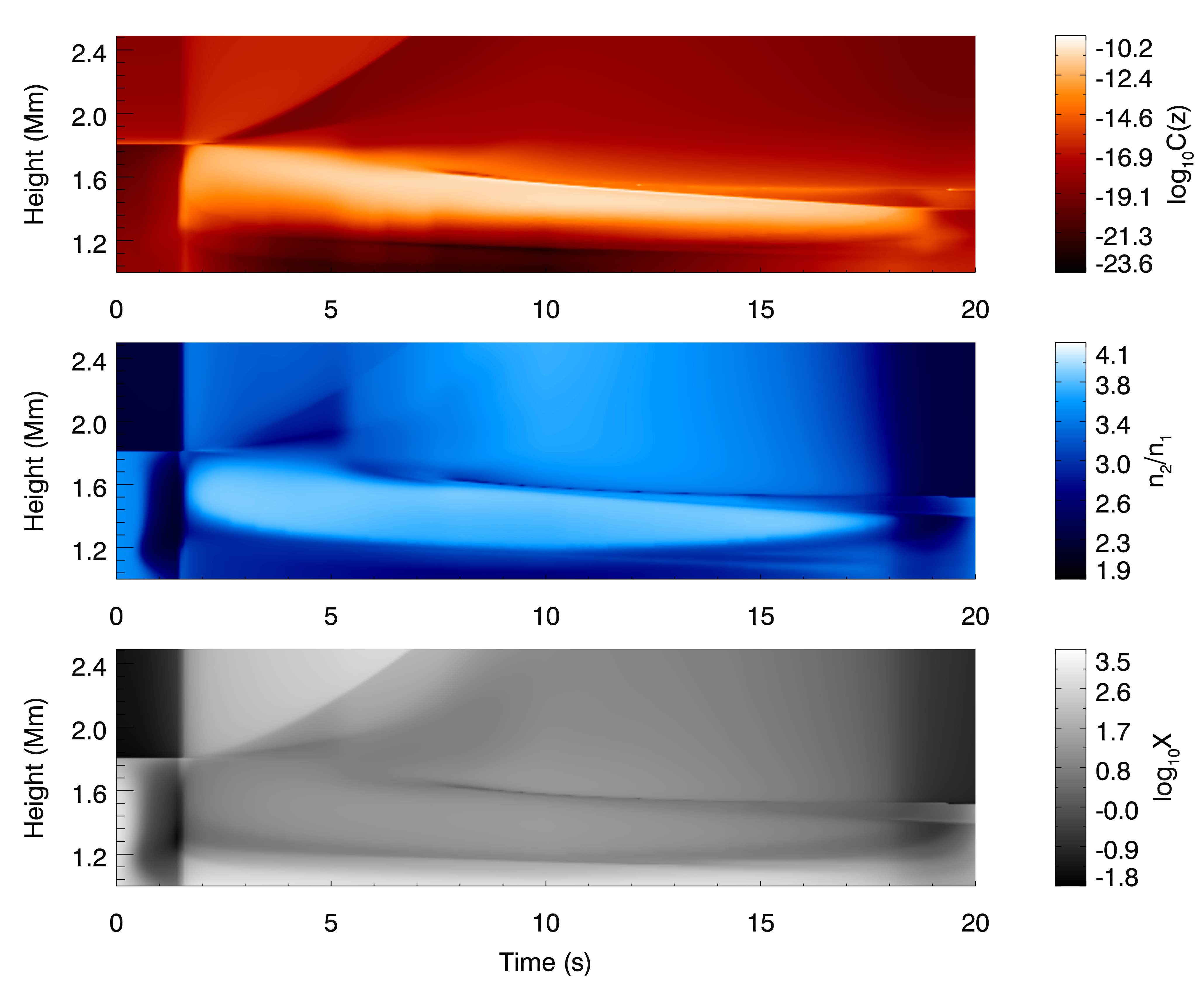}
    \caption{The same as Figure~\ref{evolution1} but for Case $\mathrm{f11E25d3}$.}
    \end{figure} 

    \begin{figure}[h]
    \centering
    \includegraphics[scale=0.48]{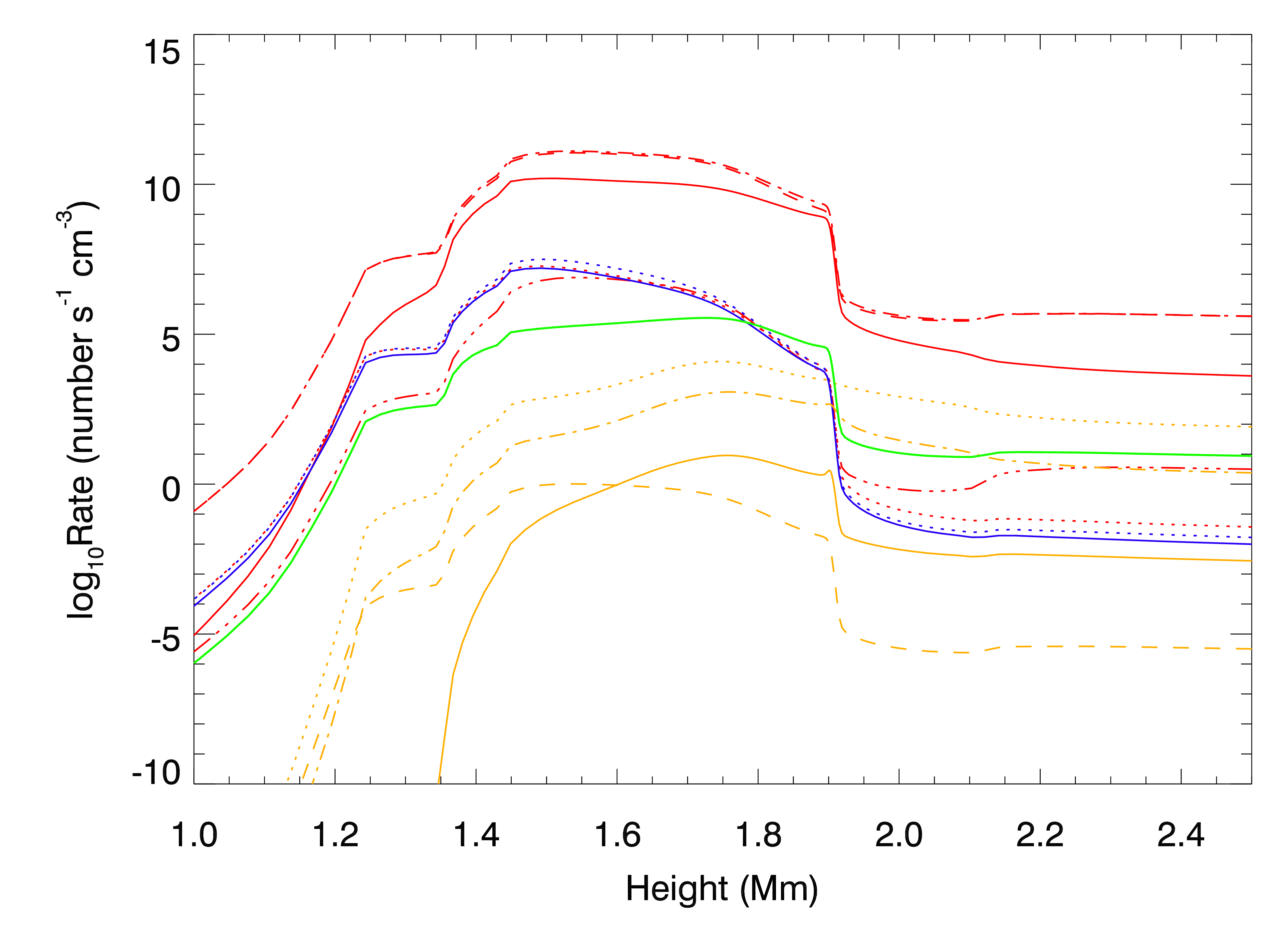}
    \caption{The same as Figure~\ref{compare} but for Case $\mathrm{f10E15d3}$.}
    \end{figure} 

    \begin{figure}[h]
    \centering
    \includegraphics[scale=0.48]{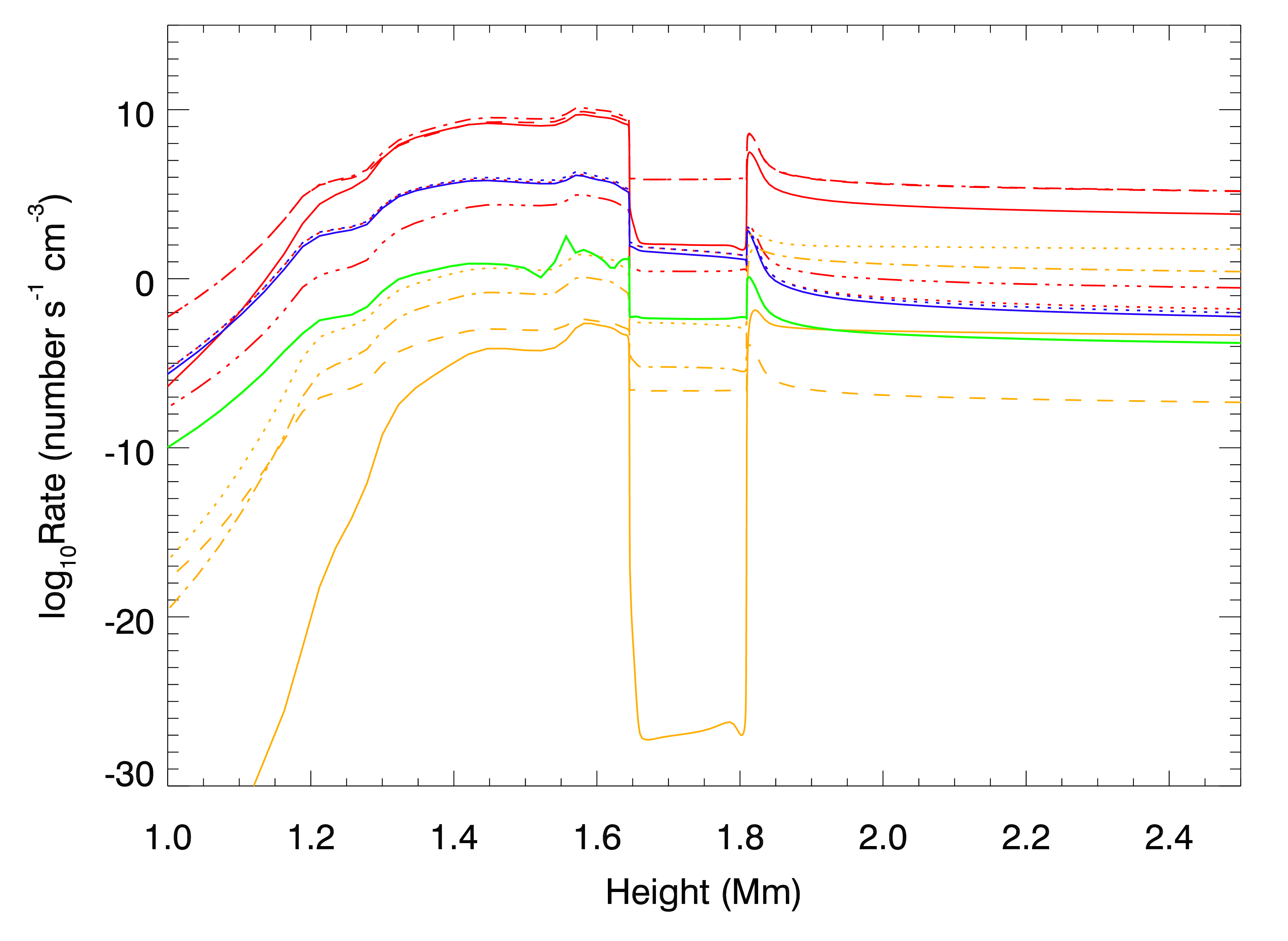}
    \caption{The same as Figure~\ref{compare} but for Case $\mathrm{f10E25d7}$.}
    \end{figure} 

    \begin{figure}[h]
    \centering
    \includegraphics[scale=0.48]{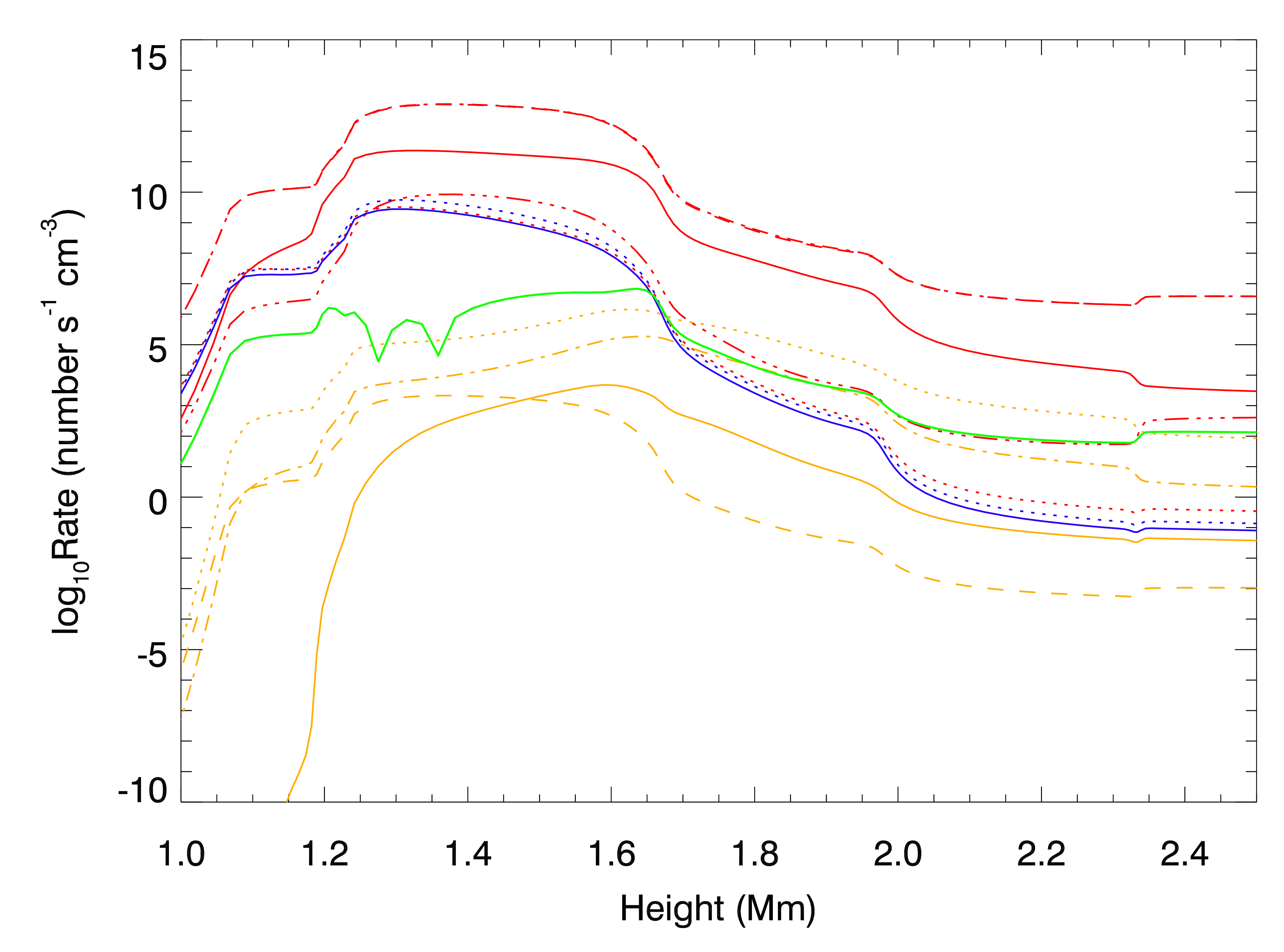}
    \caption{The same as Figure~\ref{compare} but for Case $\mathrm{f11E25d3}$.}
    \end{figure} 
    
    \begin{figure}[h]
    \centering
    \includegraphics[scale=0.48]{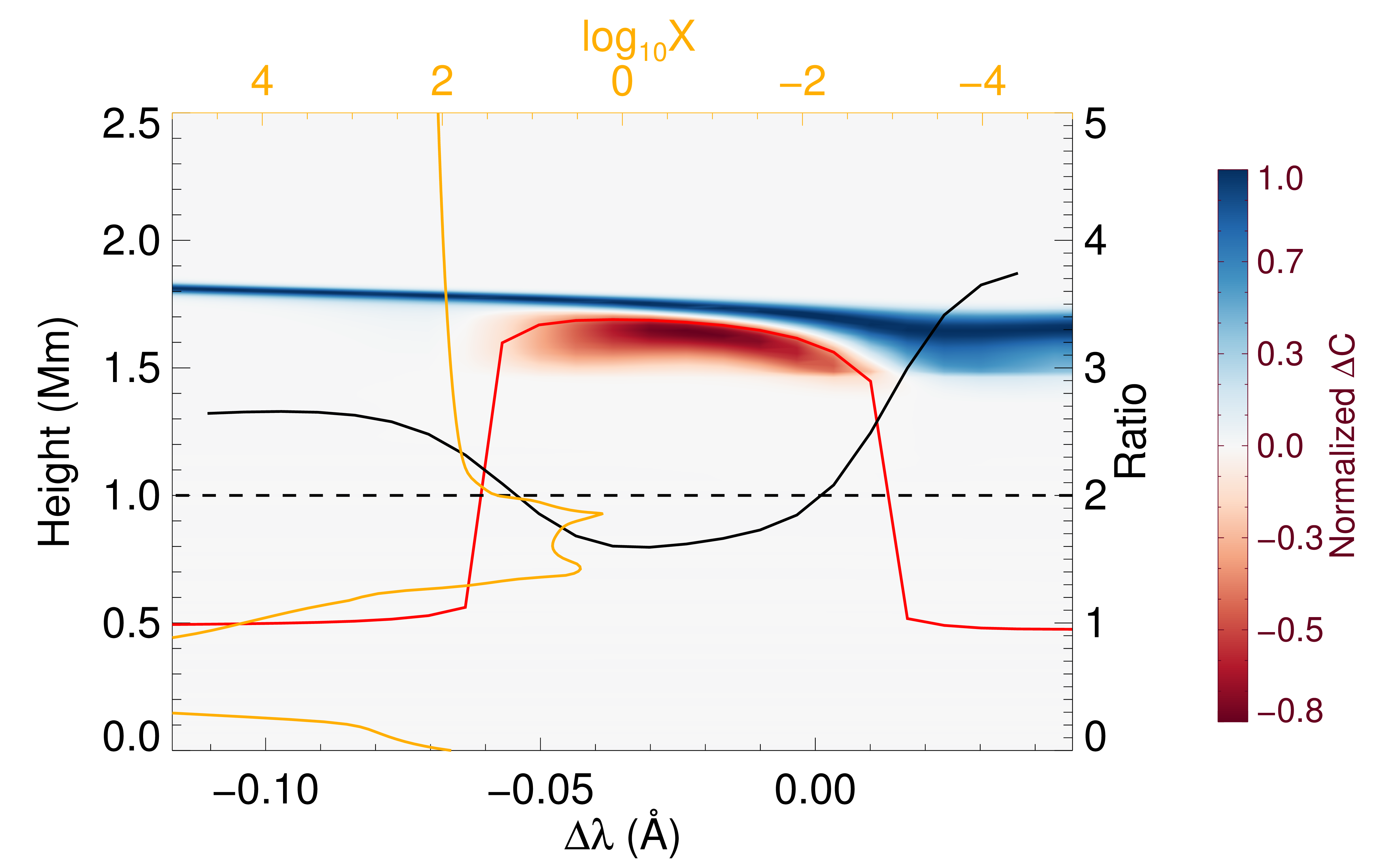}
    \caption{The same as Figure~\ref{035_RC} for Case $\mathrm{f10E15d3}$.}
    \end{figure} 

    \begin{figure}[h]
    \centering
    \includegraphics[scale=0.48]{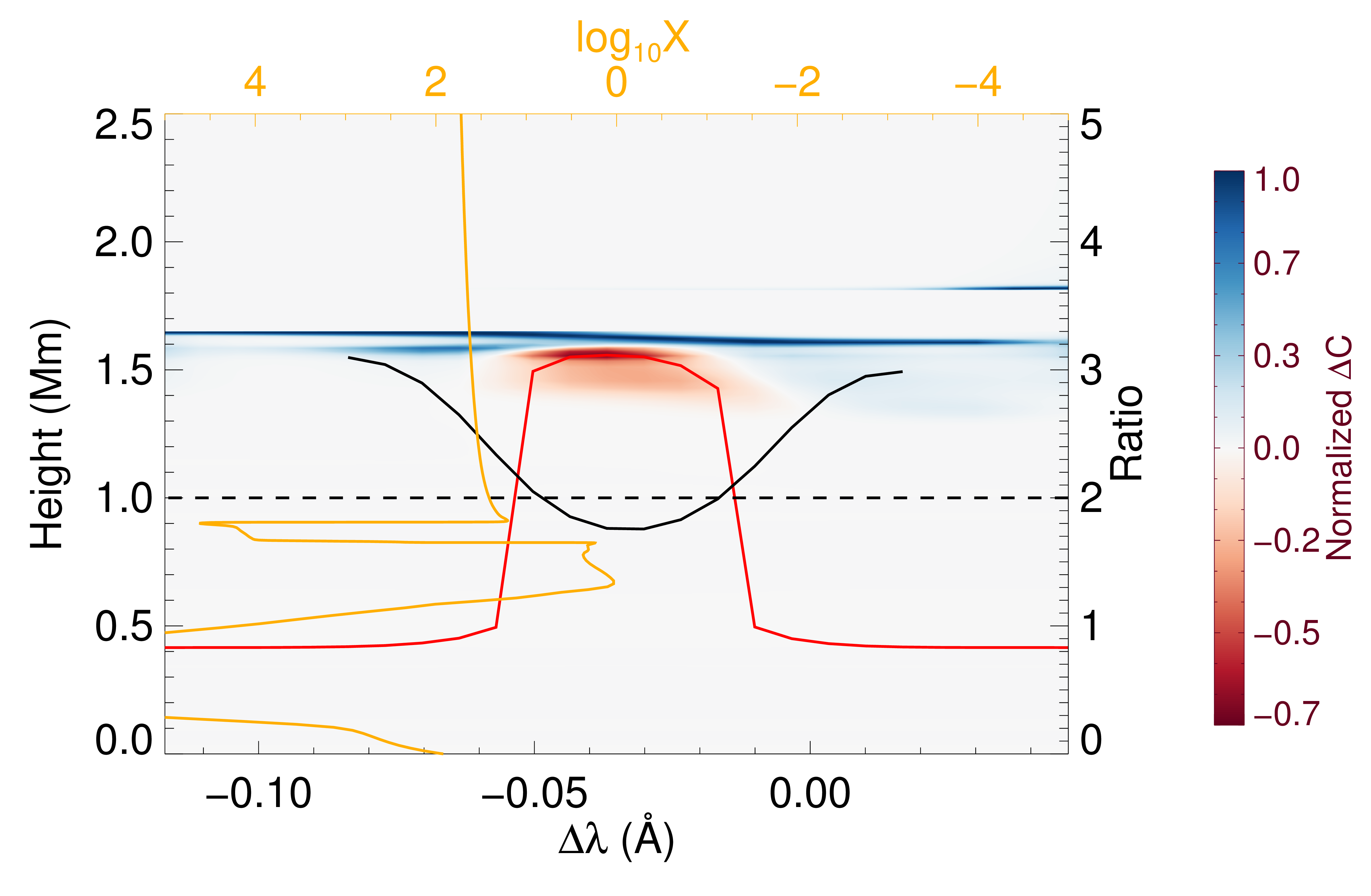}
    \caption{The same as Figure~\ref{035_RC} for Case $\mathrm{f10E25d7}$.}
    \end{figure}

    \begin{figure}[h]
    \centering
    \includegraphics[scale=0.48]{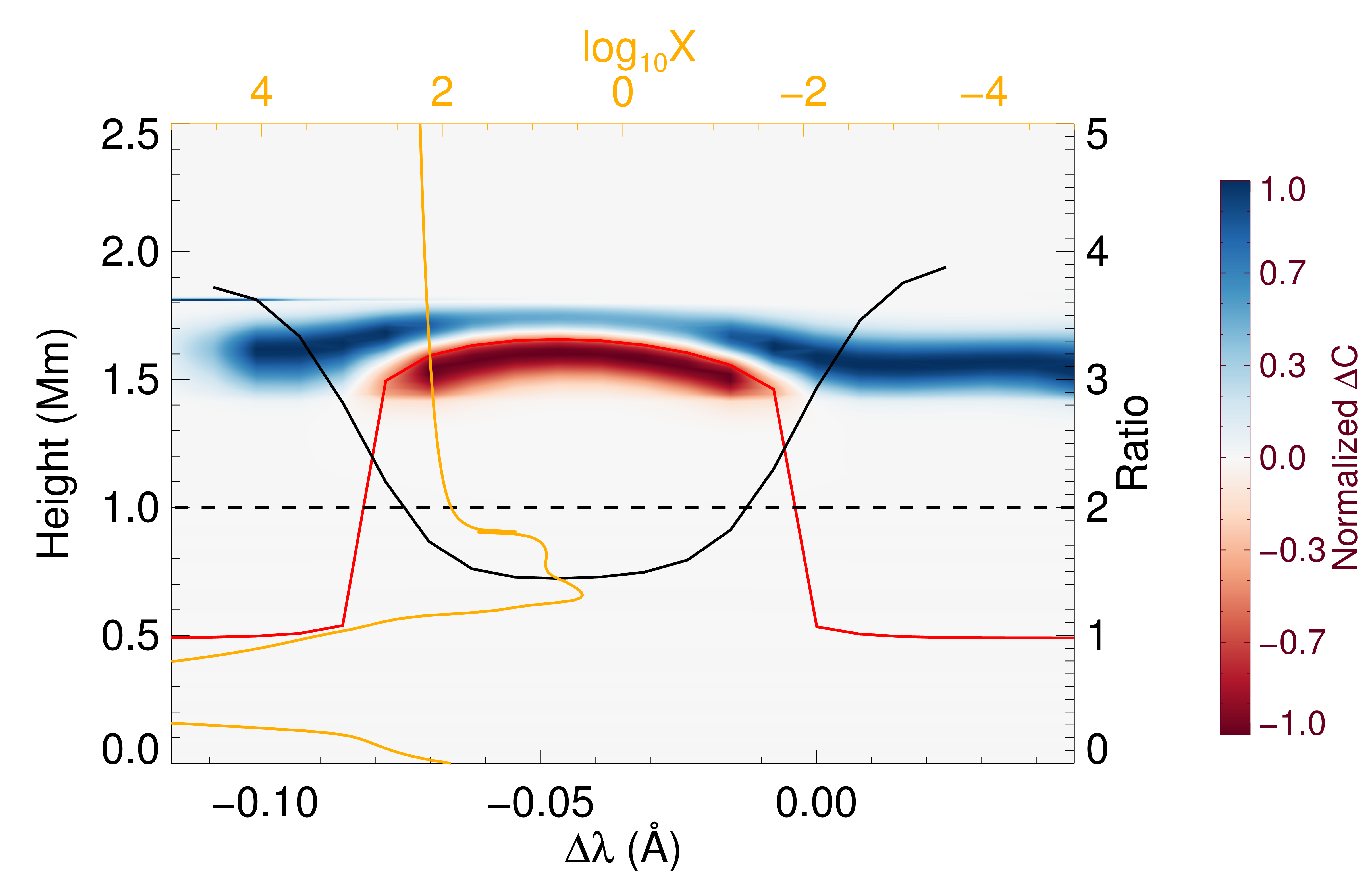}
    \caption{The same as Figure~\ref{035_RC} for Case $\mathrm{f11E25d3}$.}
    \end{figure}    
    
\end{appendix}

 \end{document}